\documentclass[a4paper,11pt]{article}
\pdfoutput=1 

\usepackage{jheppub} 
\usepackage{amsmath,amssymb,amsthm,amscd,graphicx}
\input epsf.sty

\newtheorem{theorem}{Theorem}[section]

\theoremstyle{definition}

\newtheorem{example}[theorem]{Example}

\newcommand{\CB}{{\cal B}}
\newcommand{\CC}{{\cal C}}

\newcommand{\CJ}{{\cal J}}

\newcommand{\CN}{{\cal N}}
\newcommand{\CO}{{\cal O}}

\newcommand{\CR}{{\cal R}}
\newcommand{\CS}{{\cal S}}

\def\IZ{{\mathbb Z}}
\def\IR{{\mathbb R}}
\def\IC{{\mathbb C}}
\def\IP{{\mathbb P}}

\def\IS{{\mathbb S}}
\def\IF{{\mathbb F}}
\def\IN{{\mathbb N}}

\newcommand{\tr}{\mathop{\rm Tr}\nolimits}
\newcommand{\re}{{\rm e}}
\newcommand{\ri}{{\rm i}}
\newcommand{\rd}{{\rm d}}

\newcommand{\pd}{\partial}
\def\cO{\mathcal{O}}
\def\({\left(}
\def\){\right)}

\newcommand{\be}{\begin{equation}}
\newcommand{\ee}{\end{equation}}
\newcommand{\ba}{\begin{aligned}}
\newcommand{\ea}{\end{aligned}}
\newcommand{\ben}{\begin{eqnarray}\displaystyle}
\newcommand{\een}{\end{eqnarray}}

\newcommand{\sectiono}[1]{\section{#1}\setcounter{equation}{0}}

\newdimen\tableauside\tableauside=1.0ex
\newdimen\tableaurule\tableaurule=0.4pt
\newdimen\tableaustep
\def\phantomhrule#1{\hbox{\vbox to0pt{\hrule height\tableaurule width#1\vss}}}
\def\phantomvrule#1{\vbox{\hbox to0pt{\vrule width\tableaurule height#1\hss}}}
\def\sqr{\vbox{%
  \phantomhrule\tableaustep
  \hbox{\phantomvrule\tableaustep\kern\tableaustep\phantomvrule\tableaustep}%
  \hbox{\vbox{\phantomhrule\tableauside}\kern-\tableaurule}}}
\def\squares#1{\hbox{\count0=#1\noindent\loop\sqr
  \advance\count0 by-1 \ifnum\count0>0\repeat}}
\def\tableau#1{\vcenter{\offinterlineskip
  \tableaustep=\tableauside\advance\tableaustep by-\tableaurule
  \kern\normallineskip\hbox
    {\kern\normallineskip\vbox
      {\gettableau#1 0 }%
     \kern\normallineskip\kern\tableaurule}%
  \kern\normallineskip\kern\tableaurule}}
\def\gettableau#1{\ifnum#1=0\let\next=\null\else
\squares{#1}\let\next=\gettableau\fi\next}

\newcommand{\figref}[1]{Fig.~\protect\ref{#1}}

\title{\boldmath Topological Strings from Quantum Mechanics}

\author[a]{Alba Grassi,}
\author[b]{Yasuyuki Hatsuda}
\author[a]{and Marcos Mari\~no}
\affiliation[a]{D\'epartement de Physique Th\'eorique et section de Math\'ematiques\\
Universit\'e de Gen\`eve, Gen\`eve, CH-1211 Switzerland}

\affiliation[b]{DESY Theory Group, DESY Hamburg,\\ 
Notkestrasse 85, D-22603 Hamburg, Germany }

\emailAdd{alba.grassi@unige.ch, yasuyuki.hatsuda@desy.de, marcos.marino@unige.ch} 

\preprint{
\begin{flushright}
DESY 14-181 \\
\end{flushright}
}

\abstract{We propose a general correspondence which associates a non-perturbative 
quantum-mechanical operator to a toric Calabi--Yau manifold, and we 
conjecture an explicit formula for 
its spectral determinant in terms of an M-theoretic version of the topological string free energy.
 As a consequence, we derive an exact quantization condition for the operator spectrum, in terms of the vanishing 
 of a generalized theta function. The perturbative part of this quantization condition is given by the Nekrasov--Shatashvili limit of the 
 refined topological string, but there are non-perturbative corrections determined by the conventional topological string. 
 We analyze in detail the cases of local $\IP^2$,  local $\IP^1 \times \IP^1$ and local $\IF_1$. In all these cases, the 
 predictions for the spectrum agree with the existing numerical results. We also show explicitly that 
 our conjectured spectral determinant leads to the correct spectral traces of the corresponding operators. 
 Physically, our results provide a non-perturbative formulation of topological strings on toric Calabi--Yau manifolds, in which 
 the genus expansion emerges as a 't Hooft limit of the spectral traces. Since the spectral determinant is an entire function on moduli space, it leads to a background independent 
 formulation of the theory. Mathematically, our results lead to precise, surprising conjectures 
 relating the spectral theory of functional difference operators to enumerative geometry. 
}

\begin{document}
\maketitle
\flushbottom

\sectiono{Introduction}

As it is well-known, string theory is in principle only defined perturbatively. In the last years, 
thanks to the AdS/CFT correspondence, non-perturbative formulations 
have been found in certain backgrounds, in terms of a dual gauge theory. 
The combination of this duality with localization and integrability techniques have provided us 
with concrete non-perturbative expressions for many quantities. In general, 
these quantities have a perturbative genus expansion determined by string perturbation theory, but they involve 
additional non-perturbative contributions. A particularly interesting example of such a 
quantity is the partition function of ABJM theory \cite{abjm} on the three-sphere. Using localization, 
this partition function can be expressed in terms of a matrix integral \cite{kwy}. The 't Hooft expansion of 
this integral, fully determined in \cite{dmp}, gives the genus expansion of the dual type IIA superstring. 
However, there are additional non-perturbative corrections which were first pointed out in 
\cite{dmpnp} and then uncovered in a series of papers \cite{mp,hmo, hmo2, cm,hmo3,hmmo, km}. One key idea 
in the study of the non-perturbative structure beyond the genus expansion is the formulation of the 
matrix model in terms of an ideal Fermi gas \cite{mp}, which can be in turn reduced to the spectral 
problem of an integral operator. 

The study of the ABJM matrix model indicated a close connection to topological string theory: 
its 't Hooft expansion is identical to the genus expansion of the topological string 
on the Calabi--Yau (CY) manifold known as local $\IP^1 \times \IP^1$ \cite{mpabjm,dmp}. In additon, the WKB analysis of the spectral problem of the 
Fermi gas is related to the refined topological string on the same manifold \cite{hmmo,km}, in the 
so-called Nekrasov--Shatashvili (NS) limit \cite{ns}. It is then natural to speculate that similar structures 
could be found in topological string theory on other local CY manifolds. This had been 
already pointed out in \cite{mp,mm-talk}. In \cite{hmmo} a concrete proposal was made for a 
non-perturbative topological string free energy, inspired by the results on ABJM theory. This proposal 
has two pieces: the perturbative piece is given by the standard genus expansion of the topological string, 
while the non-perturbative piece involves the refined topological string in the NS limit. 
A crucial r\^ole in the proposal was played by the HMO cancellation mechanism \cite{hmo2}, 
which guaranteed that the total free energy was smooth. 

A dual point of view on the problem has been proposed in \cite{km}, where the starting point is the spectral problem 
associated to the quantization of the mirror curve. Let $X$ a toric CY manifold, and let $\Sigma_X$ 
be the curve or Riemann surface encoding its local mirror. The equation describing this curve 
(sometimes called the spectral curve of $X$) is of the form 
\be
\label{scurve}
W_X (\re^x, \re^p)=0.
\ee
This curve can be ``quantized", and various aspects of this quantization have 
been studied over the last years, starting with \cite{adkmv}. The quantization 
of the curve promotes it to a functional difference operator, which can then be studied in the 
WKB approximation. Inspired by the work of \cite{ns}, it was found in \cite{mirmor,acdkv,nps} that 
the perturbative WKB quantization condition for the spectrum of these operators is closely related to
 the NS limit of the refined topological string on $X$. However, it was pointed out in \cite{km} that, 
if one looks at the {\it actual} spectrum of these operators, this perturbative quantization condition can not be the 
whole story, and additional non-perturbative information is needed. Moreover, \cite{km} proposed a 
non-perturbative quantization condition, based on the results of \cite{hmmo}, in which 
the perturbative result is complemented by instanton effects coming from the {\it standard} topological string. 
This condition turned out to lead to the correct spectrum 
in some special cases \cite{km,cgm}. Although \cite{km} focused on the case of local $\IP^1 \times \IP^1$, relevant for 
ABJM theory, it was suggested there that a similar story should 
apply to more general toric CY manifolds. This suggestion was pursued in
\cite{hw}, where the spectrum of the operators associated to some other toric CYs was studied numerically in full detail. 
The results of \cite{hw,fhw} indicated that, in general, the quantization condition suggested in \cite{km} 
required additional corrections. 

In this paper we will propose a detailed conjecture on the relation between 
non-perturbative quantum operators and local mirror symmetry. We will associate to each 
spectral curve (\ref{scurve}) an operator $\hat \rho_X$ with a positive, discrete spectrum, such 
that all the traces $\tr \hat \rho_X^n$, $n=1,2,\cdots$, are well-defined (technically, 
$\hat \rho_X$ is  a positive-definite, trace 
class operator.) A natural question is then: 
what is the {\it exact} spectrum of this operator? This is a sharp and concrete 
question, since as it was first noted in \cite{km} and further studied in \cite{hw}, it is possible to calculate this 
spectrum numerically. Our proposal is that the spectral determinant of $\hat \rho_X$ is encoded in the non-perturbative topological 
string free energy $J_X$ constructed in \cite{hmmo}. As we will explain, this free energy (which we will call the modified grand potential of $X$) 
defines a generalized theta function. The zeros of the spectral determinant are the zeros of this generalized theta function, and this leads to an 
exact quantization condition for the spectrum that agrees with all existing numerical results for these operators. 
In particular, the proposal of \cite{km} is a natural first approximation to our 
full quantization condition, and our conjecture explains naturally why it predicts the right spectrum 
in some special cases. In the general case, we can compute analytically the corrections to the quantization 
condition of \cite{km}, and we find that they perfectly agree with the numerical results for the spectrum found in \cite{hw}. 
The proposal we make in this paper clarifies the r\^ole of the non-perturbative free energy 
of \cite{hmmo}, and its precise relation to the exact quantization condition. 
But it also gives more information on the spectrum than just the quantization condition, since 
it provides in principle an exact expression for the spectral determinant of the corresponding operators. 
In addition, the spectral traces of the operators can be obtained from the behavior of topological string theory near the orbifold point. 

As it was already emphasized in \cite{hmmo,km}, our proposal can be regarded as a non-perturbative completion of the topological string, in which the 
topological string and the refined topological string complement each other non-perturbatively. There have been many proposals for a non-perturbative definition of the 
topological string, and in a sense this is not a well-posed problem, since there might be many different non-perturbative completions (as it happens for example 
in 2d gravity.) In fact, there is strong evidence \cite{dmp,gmz} that 
in many cases the genus expansion of the topological string is Borel summable, so one could take the Borel resummation of this series as a 
non-perturbative definition. We believe that our proposal is an interesting solution to this problem for three reasons. 

First of all, our starting point is the spectral determinant of the operator $\hat \rho_X$, which is 
well-defined and an entire function on the moduli space of $X$. This means in particular that our starting point is background independent. At the same time, different 
approximation schemes for the computation 
of this spectral determinant are encoded in different perturbative topological string amplitudes. 
For example, given the operator $\hat \rho_X$, we can define a partition function $Z_X (N, \hbar)$, 
which is well-defined for any integer $N$ and any real coupling $\hbar$. In the 't Hooft limit, 
\be
N \rightarrow \infty, \qquad {N \over \hbar} \quad  {\text{fixed}},
\ee
this partition function has a 't Hooft expansion which is determined by the standard genus expansion of the topological string on $X$.

Second, our proposal can be regarded as a concrete M-theoretic version of the topological string, 
in the spirit of the M-theory expansion of Chern--Simons--matter theories \cite{hkpt,mp}. For example, the partition function $Z_X(N, \hbar)$ has an 
M-theory expansion at large $N$ but {\it fixed} $\hbar$ which involves 
in a crucial way the Gopakumar--Vafa invariants of $X$. However, it also includes additional 
non-perturbative corrections which in particular 
cure the singularities of the Gopakumar--Vafa free energy, as in the HMO mechanism. We also have, naturally, that
\be
-\log Z_X (N, \hbar)\approx N^{3/2} \hbar^{1/2}, \quad N \gg 1, 
\ee
as in a theory of $N$ M2-branes \cite{kt}. This suggests that the physical theory underlying the spectral theory of the operator $\hat \rho_X$ 
might be a theory of M2-branes. It should be noted as well that what our proposal can be understood as a Fermi gas formulation of topological string theory, similar 
to the Fermi gas formulation of ABJM theory in \cite{mp}: the spectrum of the operator $\hat \rho_X$ gives the energy levels of the fermions, and the spectral determinant is naturally 
interpreted as the 
grand canonical partition function of this gas. 

Third, our proposal has a surprising mathematical counterpart: it leads to precise and testable predictions for the spectral determinant and the spectrum of non-trivial 
functional difference operators. According to our conjecture, the answer to these questions involves the refined BPS invariants of local CYs. In this way, we link two mathematically 
well-posed problems (the spectral theory of these operators, and the generalized enumerative geometry of CYs) in a novel way.

Although we believe that our proposal will hold for very general toric CY manifolds, in this paper we will focus for simplicity on those geometries whose mirror curve has genus one. 
In that case, the theory is simpler and we can make precision, non-trivial checks of our proposal. The details of the generalization to higher genus will be studied in a forthcoming publication. 

This paper is organized as follows. In section 2 we present the correspondence between mirror curves and quantum operators. 
In section 3 we state our conjecture for the spectral determinant of these operators, we derive the quantization condition implied by our 
conjecture, we comment on the physical implications of our results, and we study the simplest cases of our theory, which we call the ``maximally supersymmetric cases." Section 4 
presents a detailed illustration of our claims in the case of local $\IP^2$. Section 5 presents additional evidence for our conjecture 
by looking at two other geometries: local $\IF_1$, and local $\IP^1 \times \IP^1$, which was the original testing ground due to its relationship to ABJM theory. Finally, in section 
6 we conclude and list various open problems. 
In appendix~A, we give a derivation of the first quantum correction to the grand potential of local $\IP^2$.

 \sectiono{From mirror curves to quantum operators}

In this section we will present a correspondence between mirror curves and quantum operators. Aspects of this correspondence have been 
explored in various papers, starting in \cite{adkmv} and, more relevant to our purposes, in \cite{mirmor,acdkv}, building on the work of \cite{ns} for 
gauge theories. However, our interest will be in defining a {\it non-perturbative} spectral problem, from which one can compute a well-defined spectrum. This was first 
proposed in \cite{km} and then pursued in \cite{hw}. 

Let us start by reminding some basic notions of local mirror symmetry \cite{kkv,ckyz}. 
We consider the A-model topological string on a (non-compact) toric CY threefold, which can be 
described as a symplectic quotient 
\be
X=\mathbb{C}^{k+3}//G,
\ee
where $G=U(1)^k$. Alternatively, $X$ may be viewed physically as 
the moduli space of vacua for the complex scalars $\phi_i$, $i=0,\ldots,k+2$
of chiral superfields in a 2d gauged linear, $(2,2)$ supersymmetric 
$\sigma$-model~\cite{witten-phases}. These fields transform as 
\be
\phi_i\rightarrow \re^{ \ri Q_i^\alpha  \theta_\alpha} \phi_i, \qquad  Q_i^\alpha\in \mathbb{Z}, \quad \alpha=1,\ldots, 
k
\ee
under the gauge group $U(1)^k$. Therefore, $X$ is determined by the $D$-term constraints
\begin{equation} 
\sum_{i=0}^{k+2} Q_i^\alpha|X_i|^2=r^\alpha, \quad \alpha=1,\ldots, k 
\label{dterm} 
\end{equation}
modulo the action of $G=U(1)^k$. The $r^\alpha$ correspond to the K\"ahler parameters. The CY condition 
$c_1(TX)=0$ holds if and only if the charges satisfy \cite{witten-phases}
\begin{equation}  
\sum_{i=0}^{k+2} Q_i^\alpha=0, \qquad \alpha = 1, \ldots, k .
\label{anomaly}
\end{equation}  

The mirrors to these toric CYs were constructed by~\cite{hv}, extending~\cite{kkv, Bat}. They involve 
$3+k$ dual fields $Y^i$, $i=0, \cdots, k+2$, living in $\IC^*$. The D-term equation 
(\ref{dterm}) leads to the constraint
\be
\label{dterm-Y}
\sum_{i=0}^{k+2} Q_i^\alpha Y^i= \log z_\alpha,  \qquad \alpha = 1, \ldots, k. 
\ee
Here, the $z_{\alpha}$ are moduli parametrizing the complex structures of the mirror $\widehat X$, which is given by 
\be
w^+w^-= W_X ,
\ee
where
\be
\label{wx}
W_X= \sum_{i=0}^{k+2} \re^{Y_i}. 
\ee
The constraints (\ref{dterm-Y}) have a three-dimensional family of solutions. One of the parameters correspond to a translation of all the fields 
\be
\label{trans}
Y^i \rightarrow Y^i+c, \qquad i=0, \cdots, k+2, 
\ee
which can be used for example to set one of the $Y^i$s to zero. The remaining fields can be expressed in terms of two variables which we will denote by 
$x$, $p$. The resulting parametrization has a group of symmetries given by transformations of the form \cite{akv}, 
\be
\label{can-t}
\begin{pmatrix} x \\ p \end{pmatrix}\rightarrow G \begin{pmatrix} x \\ p \end{pmatrix}, \qquad G\in {\rm SL}(2, \IZ). 
\ee
After solving for the variables $Y^i$ in terms of the variables $x$, $p$, one finds a function 
\be
W_X (\re^x, \re^p). 
\ee
Note that, due to the translation invariance (\ref{trans}) and the symmetry (\ref{can-t}), the function 
$W_X(\re^x, \re^p)$ in (\ref{riemann}) is only well-defined up to an overall factor of the form $\re^{ \lambda x + \mu p}$, $\lambda, \mu \in \IZ$, and a transformation of the form (\ref{can-t}). 
It turns out \cite{mmopen,bkmp} that all the perturbative information about the B-model topological string on $\widehat X$ is encoded in the equation 
\be
\label{riemann}
W_X (\re^x, \re^p)=0, 
\ee
which can be regarded as the equation for a Riemann surface $\Sigma_X$ embedded in $\IC^* \times \IC^*$. 

In this paper we will focus for simplicity on toric CY manifolds $X$ in which $\Sigma_X$ has genus one, i.e. it is an 
elliptic curve\footnote{When $\Sigma_X$ has genus zero, the operator associated 
to $X$ does not seem to have a discrete spectrum, therefore we will not consider this case in this paper.}. 
The most general class of such manifolds are 
toric del Pezzo CYs, which are 
defined as the total space of the canonical bundle on a del Pezzo surface\footnote{Sometimes a distinction is made between del Pezzo surfaces and almost del Pezzo surfaces. Since our results apply to both of them, we will 
call them simply del Pezzo surfaces.} $S$,
\be
\label{dP}
\CO(K_S) \rightarrow S. 
\ee
These manifolds can be classified by reflexive polyhedra in two dimensions (see for example \cite{ckyz,hkp} for a review of this and other facts on these 
geometries). The polyhedron $\Delta_S$ associated to a surface $S$ is the convex hull of a set of two-dimensional 
vectors 
\be
\nu^{(i)}=\left(\nu^{(i)}_1, \nu^{(i)}_2\right), \qquad i=1, \cdots, k+2. 
\ee
The extended vectors 
\be
\ba
\overline \nu^{(0)}&=(1, 0,0), \\
\overline \nu^{(i)}&=\left(1, \nu^{(i)}_1, \nu^{(i)}_2\right),\qquad i=1, \cdots, k+2,
\ea
\ee
satisfy the relations 
\be
\sum_{i=0}^{k+2} Q^\alpha_i \overline \nu^{(i)}=0, 
\ee
where $Q^\alpha_i$ is the vector of charges characterizing the geometry in (\ref{dterm}).
Note that the two-dimensional vectors $\nu^{(i)}$ satisfy,  
\be
\label{vanish}
\sum_{i=1}^{k+2} Q^\alpha_i\nu^{(i)}=0. 
\ee
It turns out that the complex moduli of the mirror $\widehat X$ 
are of two types: one of them, which we will denote $\tilde u$ as in \cite{hkp,hkrs}, 
is a ``true" complex modulus for the elliptic curve $\Sigma$, and it is associated to the compact four-cycle $S$ in $X$. The remaining moduli, which will be denoted as $m_i$, 
should be regarded as parameters. For local del Pezzos, there is a canonical parametrization of the curve (\ref{riemann}), as follows. Let 
\be
\ba
Y^0&=\log \tilde u, \\
Y^i & =\nu^{(i)}_1 x+  \nu^{(i)}_2 p + f_i(m_j), \qquad i=1, \cdots, k+2. 
\ea
\ee
Due to (\ref{vanish}), the terms in $x$, $p$ cancel, as required to satisfy (\ref{dterm-Y}). In addition, we find the parametrization 
\be
\log z_\alpha=  \log \tilde u^{Q^\alpha_0}+ \sum_{i=1}^{k+2} Q^\alpha_i f_i(m_j), 
\ee
which can be used to solve for the functions $f_i(m_j)$, up to reparametrizations. We then find the equation for the curve, 
\be
\label{ex-W}
W_X= \CO_X(x,p)+ \tilde u=0,  
\ee
where
\be
\label{coxp}
 \CO_X(x,p)=\sum_{i=1}^{k+2} \exp\left( \nu^{(i)}_1 x+  \nu^{(i)}_2 p + f_i(m_j) \right). 
 \ee
 %

Let $\hat x$ and $\hat p$ be standard quantum-mechanical operators 
satisfying the canonical commutation relation
\be
[\hat x, \hat p]= \ri \hbar. 
\ee
In this paper, $\hbar$ will be a real parameter. We need to consider as well the exponentiated operators
\be
\hat X= \re^{\hat x} , \qquad \hat P=\re^{\hat p}. 
\ee
These operators are self-adjoint and they satisfy the Weyl algebra
\be
\hat X \hat P= q  \hat P \hat X, 
\ee
where
\be
\label{q-def} q=\re^{\ri \hbar}. 
\ee
However, the domains of $X$, $P$ should be defined appropriately, since they lead to difference or displacement operators 
acting on wavefunctions (for example, if we work in the 
$x$ representation, $P$ is a difference operator.) The domain of the operator $X$, $D(X)$, consists of wavefunctions $\psi(x) \in L^2(\IR)$ such that
\be
\re^x \psi(x) \in L^2(\IR).  
\ee
Similarly, the domain of $P$, $D(P)$, consists of functions $\psi(x) \in L^2(\IR)$ such that 
\be
\label{p-cond}
\re^p \widehat \psi(p) \in L^2(\IR), 
\ee
where 
\be
\widehat \psi(p)=\int {\rd x \over {\sqrt{2 \pi \hbar}}} \re^{-\ri p q} \psi(q)
\ee
is the wavefunction in the $p$ representation, which is essentially given by a Fourier transform. The condition (\ref{p-cond}) can be translated 
into a condition on $\psi(x)$ (see for example \cite{ft}): this is a function 
which admits an analytic continuation into the strip 
\be
\CS_{-\hbar}= \left\{ x- \ri y \in \IC: 0< y< \hbar\right\}, 
\ee
such that $\psi(x-\ri y) \in L^2(\IR)$ for all $0\le y<\hbar$, and the limit
\be
\psi\left(x-\ri \hbar+\ri 0 \right) =\lim_{\epsilon \to 0^+} \psi \left(x- \ri \hbar+\ri \epsilon \right)
\ee
exists in the sense of convergence in $L^2(\IR)$. 

We want now to associate a self-adjoint quantum operator $\widehat \CO_X $ of the form 
\be
\label{oxop}
\widehat \CO_X (\hat x, \hat p)=\sum_{r,s \in \IZ } a_{r,s} \re^{r \hat x+ s \hat p}, \qquad a_{r,s}\ge 0,
\ee
 to each toric del Pezzo $X$, in such a way that we have a well-defined eigenvalue problem
\be
\label{spec}
\widehat \CO_X (\hat x, \hat p) |\psi_n \rangle =\re^{E_n} |\psi_n \rangle, \qquad n=0,1,\cdots, 
\ee
i.e. we want to have a {\it discrete} and {\it positive} spectrum, so that the energies $E_n$ are real. 
It is convenient to consider the inverse operator 
\be
\hat \rho_X=\widehat \CO_X^{-1} \left(\hat x, \hat p\right). 
\ee
The {\it spectral traces} of $\hat \rho_X$ are defined by  
\be
\label{spec-tr}
Z_\ell= \tr \hat \rho_X^{\ell}=\sum_{n=0}^\infty \re^{-\ell E_n}, \qquad \ell=1,2,\cdots, 
\ee
and we will require them to be well-defined (i.e. finite). The semiclassical limit of these traces is given by, 
\be
Z_\ell \approx {1\over \hbar} Z^{(0)}_\ell , \qquad \hbar \rightarrow 0, 
\ee
where
\be
\label{cl-traces}
Z^{(0)}_\ell = \int {\rd x \rd p \over 2 \pi } {1\over \left( \CO_X(x, p) \right)^{\ell}}, 
\ee
and $\CO_X(x,p)$ denotes the classical function underlying (\ref{oxop}), 
or more formally, the Wigner transform of the operator (\ref{oxop}) (this 
classical function is simply given by the expression (\ref{oxop}) where 
we replace $\hat x$, $\hat p$ by the corresponding classical variables.) 
If the semiclassical limit is smooth, as we will assume here, we should have
\be
\label{finite-trace}
Z_\ell^{(0)}<\infty.
\ee
This leads to useful constraints on the form of $\CO_X(x,p)$.

Let us explain how to associate a quantum operator to a given local del Pezzo. 
We have seen in (\ref{ex-W}) that, for local del Pezzo's, the function $W_X(\re^x, \re^p)$ can always be written in the form (\ref{ex-W}).
The operator $\widehat \CO_X(\hat x, \hat p)$ is obtained by 
promoting the classical function $\CO_X(x, p)$ in (\ref{coxp}) to a quantum operator. In this promotion, we use Weyl's prescription for ordering ambiguities. 
This associates
\be
\re^{r x + s p} \rightarrow  \re^{r \hat x + s \hat p}, 
\ee
so that the resulting operator is Hermitian. Clearly, if the parameters $m_i$ satisfy appropriate reality and positivity conditions, the resulting quantum operator will be of 
the form (\ref{oxop}). Since this operator is a sum of operators of the form $ \re^{r \hat x + s \hat p}$, its domain is given by the intersection of the domains of all the operators of this type 
appearing in the sum in (\ref{oxop}). 

\begin{table}
\centering
\begin{tabular}{||  l || l || l || l ||}
\hline
$X$  & $\CO_X (x,p)$ & $C$ & $r$ \\ \hline\hline
local $\IP^2$ &  $\re^{x}+ \re^p + \re^{-x-p}$ & $9/2$ & $3$ \\ \hline \hline
local $\IF_0$ &  $\re^x+ m \re^{-x} + \re^p + \re^{-p}$ & $4$ & $2$ \\ \hline \hline
local $\IF_1$ &  $\re^x+ m \re^{-x} + \re^p +  \re^{-x -p}$& $ 4 $ & $1$  \\ \hline \hline
local $\IF_2$ & $\re^x+ m \re^{-x} + \re^p + \re^{ -2x-p }$& $ 4 $  & $2$ \\ \hline \hline
local $\CB_2$ & $m_2 \re^x+m_1 \re^p +  \re^{-x} +  \re^{-p } +  \re^{x+p}$& $ 7/2$  & $1$ \\ \hline \hline
local $\CB_3$ & $m_1 \re^{-x} + \re^x + m_2 \re^{-p}+ \re^p + m_3 \re^{x+ p}+ \re^{-x-p}$& $ 3$ & $1$ \\ \hline 
\end{tabular}
\caption{In this table we list the operators associated to some local del Pezzo CYs, as well as the values of the constant $C$ defined by (\ref{as-vol}) and the index $r$ by (\ref{z-mod}).}
\label{table-ops}
\end{table}

\begin{example} In order to illustrate this procedure, let us consider the well-known example of local $\IP^2$. In this case, we have $k=1$ and 
the toric CY is defined by a single charge vector $Q=(-3,1,1,1)$. The corresponding 
polyhedron $\Delta_S$ for $S=\IP^2$ is obtained as the convex hull of the vectors
\be
\nu^{(1)}=(1,0), \qquad \nu^{(2)}=(0,1), \qquad \nu^{(3)}=(-1,-1). 
\ee
In the mirror, the variables 
$Y^i$ satisfy
\be
-3 Y^0+Y^1+ Y^2+ Y^3 =-3 \log \tilde u, 
\ee
and the canonical parametrization is given by 
\be
Y^0=\log \tilde u, \qquad Y^1=x, \qquad Y^2=p, \qquad Y^3=-x-p, 
\ee
so that
\be
W_X(\re^x, \re^p)=\re^x+ \re^p + \re^{-x-p}+ \tilde u, 
\ee
after changing $\tilde u \rightarrow -\tilde u$. Therefore, the quantum operator is given by 
\be
\label{lp2-op}
\widehat \CO_X \left(\hat x, \hat p\right)= \re^{\hat x} + \re^{\hat p} + \re^{-\hat x-\hat p}. 
\ee
This operator was studied, from a semiclassical point of view, in \cite{gkm}. Its spectrum was studied numerically in \cite{hw}. \qed \end{example} 

%
%

Following the procedure in the previous example, we can write down operators for other local del Pezzo CYs. A list with some useful examples can be found in table \ref{table-ops}, where we used for 
convenience the classical version $\CO_X(x,p)$. The conventions for the parametrization of the curves (in particular, for the parameters $m$, $m_i$ appearing 
in the equations) are those of \cite{hkrs,hkp}. Note that a transformation of the form (\ref{can-t}) corresponds to a canonical 
transformation, and will not change the spectrum of the operator. Note as well that, after changing $\tilde u \rightarrow -\tilde u$, the spectral problem (\ref{spec}) can be written as 
\be
\label{spec2}
W_X\left (\re^{\hat x}, \re^{\hat p} \right) |\psi_n \rangle=0, 
\ee
where we use the form (\ref{ex-W}). The spectral problem leads then to a quantization of the modulus $\tilde u$, 
which after the change of sign above, can be interpreted as the exponential of the energy: 
\be
\label{ure}
\tilde u=\re^E. 
\ee
We can regard $\widehat \CO_X (\re^{\hat x}, \re^{\hat p})$ as the exponential of a Hamiltonian $\hat H_X$, while $\hat \rho_X$ can be interpreted 
as the canonical density matrix, 
\be
\label{hamiltonian}
\widehat \CO_X (\re^{\hat x}, \re^{\hat p})=\re^{\hat H_X}, \qquad \hat \rho_X=\re^{-\hat H_X}. 
\ee
The operator $\hat H$ has a complicated Wigner transform (as in the closely related examples of \cite{mp}). Its explicit form will not be 
needed in this paper, but it might be useful to test some of our statements in a semiclassical analysis, as in \cite{mp}. 

\begin{figure}
\begin{center}
\begin{minipage}{0.4\hsize}
 \begin{center}
  \includegraphics[height=5cm]{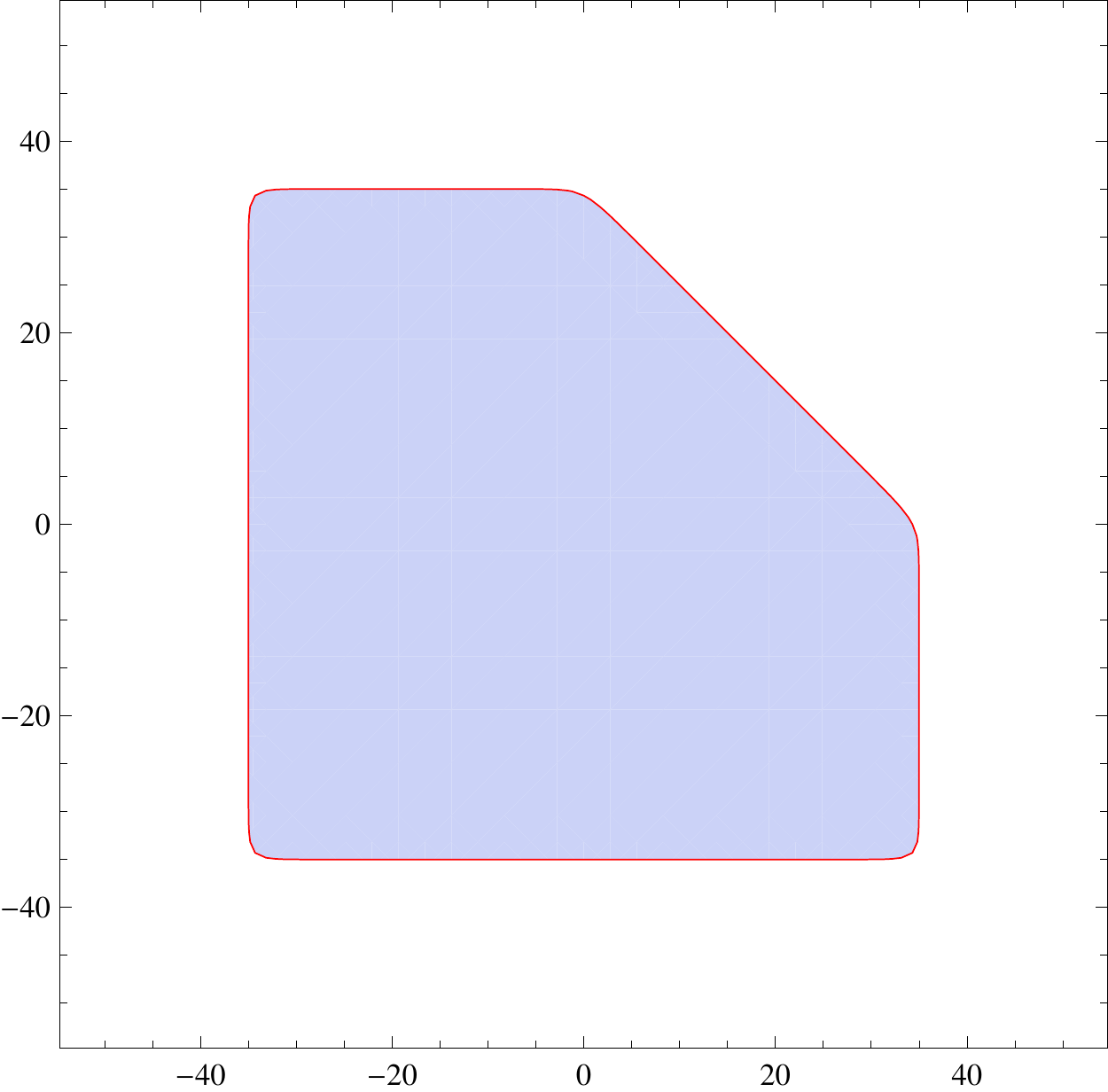}
 \end{center}
\end{minipage}
\begin{minipage}{0.4\hsize}
 \begin{center}
  \includegraphics[height=3cm]{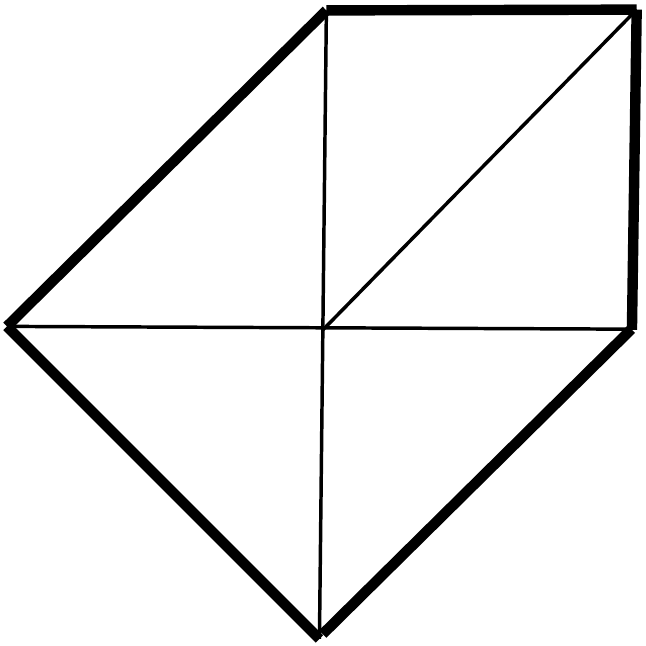}
 \end{center}
 \end{minipage}
\end{center}
\caption{The figure on the left shows the region (\ref{reg-E}) in phase space for the quantum operator associated to local $\CB_2$, for $E=35$ and $m_1=m_2=1$. The figure on the right is 
the polyhedron representing toric $\CB_2$.}
\label{b2}
\end{figure}

\begin{figure}
\begin{center}
\begin{minipage}{0.4\hsize}
 \begin{center}
  \includegraphics[height=5cm]{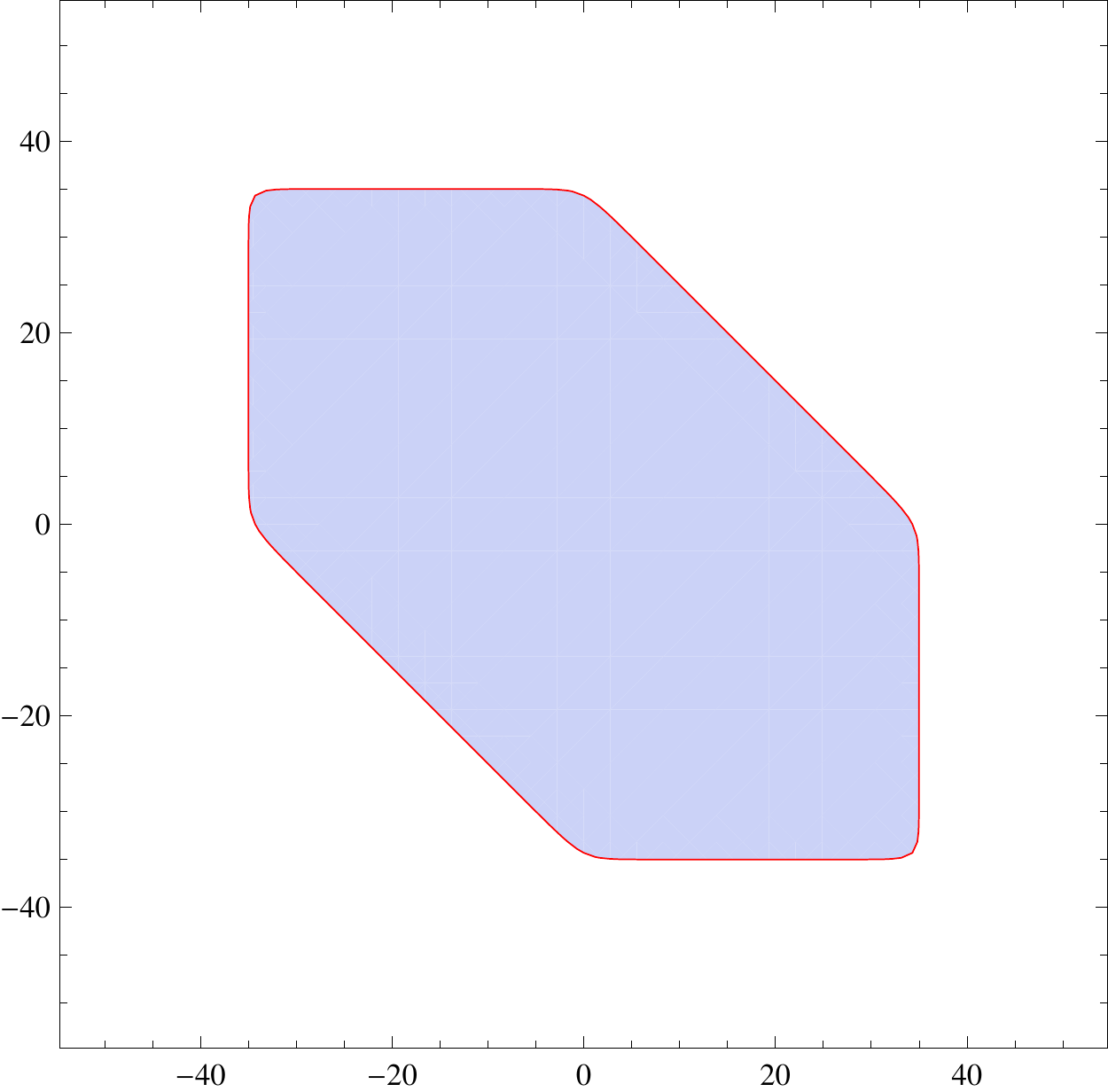}
 \end{center}
\end{minipage}
\begin{minipage}{0.4\hsize}
 \begin{center}
  \includegraphics[height=3cm]{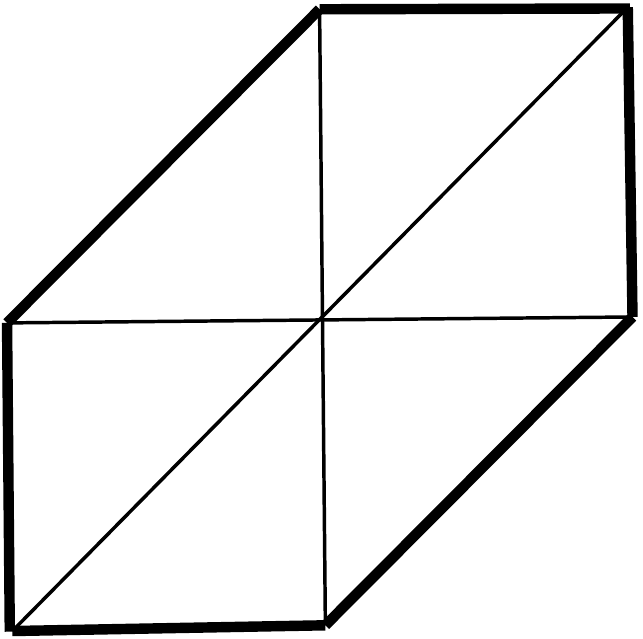}
 \end{center}
 \end{minipage}
\end{center}
\caption{The figure on the left shows the region (\ref{reg-E}) in phase space for the quantum operator associated to local $\CB_3$, for $E=35$ and $m_1=m_2=m_3=1$. The figure on the right is 
the polyhedron representing toric $\CB_3$.}
\label{b3}
\end{figure}

In order to gain some insight into these operators, and to verify that the requirement (\ref{finite-trace}) holds for them,  
we can consider their semiclassical limit and the corresponding Bohr--Sommerfeld quantization condition. The region of phase 
space with energy less or equal than $E$ is defined by the equation, 
\be
\label{reg-E}
\CR(E)=\{ (x,p) \in \IR^2: \CO_X(x,p) \le \re^E \}. 
\ee
As is well-known, in the semiclassical limit each cell of volume $2 \pi \hbar$ in $\CR(E)$ will lead to a quantum state. Therefore, 
if we want the spectrum of $\widehat \CO_X$ to be discrete, 
we should require $\CR(E)$ to have a finite volume. The geometry of the region $\CR(E)$ at large energies 
is easy to understand (and very similar to the situations considered in \cite{mp,cmp}): 
for large $E$, we should consider the tropical limit of the curve (\ref{coxp}), which in the canonical parametrization (\ref{ex-W}) reads
\be
\label{lines}
  \nu^{(i)}_1 x+  \nu^{(i)}_2 p + f_i(m_j)=E, \qquad i=1, \cdots, k+2. 
  \ee
The boundary of the 
region $\CR(E)$ is the polygon limited by the lines (\ref{lines}). This polygon is nothing but the boundary of the dual polyhedron $\Delta_S^\star$ 
defining the toric del Pezzo, see for example \figref{b2} and \figref{b3} for nice illustrations 
involving local $\CB_2$ and local $\CB_3$, respectively. Therefore, the region (\ref{reg-E}) has a finite volume. This also guarantees that the classical function 
\be
\rho_X(x,p)={1\over \CO_X(x,p)} 
\ee
decays exponentially at infinity, so that (\ref{finite-trace}) is verified.

We expect the difference operators $\widehat \CO_X(\hat x, \hat p)$ constructed in this way to have a positive and discrete spectrum. Specifically, 
we expect their inverses $\hat \rho_X$ to be positive-definite and trace class operators. This is clearly indicated by the 
behavior of the semiclassical limit, but it would be important to prove it from first principles, in order 
to make sure that the spectral problem and the spectral traces are defined rigorously\footnote{A rigorous proof for some special cases, like the operator for local $\IP^2$, appears in \cite{kas-mar}.}. 

In practice, one can calculate the spectrum of the operators $\widehat \CO_X(\hat x, \hat p)$ as in \cite{hw}\footnote{In the case of ABJM theory, it is possible to obtain an explicit form for the 
integral kernel of $\hat \rho$, and one can use standard techniques for the computation of the eigenvalues and eigenfunctions of such kernels, see \cite{hmo,km}.}: one chooses a 
system of orthonormal wavefunctions $|\varphi_n \rangle$ which belongs to $D(X)\cap D(P)$. A useful choice is the basis of eigenfunctions of the harmonic oscillator, since they have 
Gaussian decay along all parallel directions to the real axis in 
the complex plane. Then, the infinite-dimensional matrix 
\be
\label{omatrix}
\left( \widehat \CO_X  \right)_{nm}=\langle \varphi_n | \widehat \CO_X (\hat x, \hat p) |\varphi_m \rangle
\ee
can be diagonalized numerically: one first truncates it to an $L \times L$ dimensional matrix, computes the eigenvalues $E^{(L)}_n$, $n=0, 1, \cdots$, and observes numerical convergence as $L$ grows, 
\be
E_n^{(L)} \rightarrow E_n, \qquad L \rightarrow \infty, \quad n=0, 1, \cdots. 
\ee
In this paper we will rely on this method to check our analytical results on the spectrum. Detailed numerical results for the spectrum of the first two operators in table \ref{table-ops} can be found in \cite{hw}. 

\sectiono{Spectral determinants and topological strings}

In this section we state our main conjecture, which gives a conjectural expression for the 
spectral determinant of the operator $\hat \rho_X$ introduced in the previous section. We also discuss the quantization condition for the 
spectrum derived from our conjecture, as well as its physical meaning. 

\subsection{The spectral determinant}

The spectral information about the operators $\hat \rho_X$ and $\widehat \CO_X$ can be encoded in various useful ways. Given a trace class operator 
$\hat \rho$ with eigenvalues $\re^{-E_n}$, $n=0, 1, \cdots$, and depending on a real 
parameter $\hbar$, its {\it spectral determinant} (also 
called Fredholm determinant) is defined by 
\be
\label{f-det}
\Xi(\kappa, \hbar)= {\rm det}(1+ \kappa \hat \rho)= \prod_{n=0}^\infty \left(1+ \kappa\re^{-E_n} \right). 
\ee
We will refer to $\kappa$ as the fugacity, and we will often write it as
\be
\label{kappa-mu}
\kappa=\re^{\mu}, 
\ee
where $\mu$ is called the chemical potential. We will use the arguments $\kappa$ and $\mu$ interchangeably. The reason for 
this terminology is that $\Xi(\kappa, \hbar)$ can be physically interpreted as the grand canonical 
partition function of an ideal Fermi gas where the one-particle problem has energy levels $E_n$. Note that our spectral determinant is 
different from the one usually studied in Quantum Mechanics \cite{voros,dt,cvitanovic}:
\be
\label{qm-det}
D(\mu,\hbar)=\prod_{n\ge 0} \left( 1+{\mu\over E_n} \right).
\ee
Our definition (\ref{f-det}) uses instead the 
canonical density matrix. It has better convergence properties and does not need to be regularized, in contrast to (\ref{qm-det}). For example, in the case of the 
quantum harmonic oscillator, the spectral determinant (\ref{qm-det}) leads, after regularization, to 
\be
D(\mu,\hbar)={\sqrt{\pi} \over \Gamma(1/2+\mu/\hbar)}, 
\ee
while with our definition we would obtain 
\be
\Xi(\kappa, \hbar)= \prod_{n=0}^\infty \left(1+ \kappa \re^{-\hbar(n+1/2)} \right)= \left(-\re^{-\hbar/2} \kappa; \re^{-\hbar} \right)_{\infty}, 
\ee
which is the quantum dilogarithm \cite{fk}. 

The spectral determinant 
has two important properties: first, it is an {\it entire} function of the fugacity $\kappa$ (see for example \cite{simon}, chapter 3, for a proof of this fact). Second, after setting 
\be
\label{kappa-E}
\kappa=-\re^E,
\ee
it has simple zeros, as a function of $E$, at the energies of the spectrum $E_n$. This means that one can in principle 
read the spectrum of the operator $\hat \rho$ by looking at the zeros of the spectral determinant. The grand potential is defined as
\be
\CJ(\mu, \hbar)= \log \Xi(\mu, \hbar), 
\ee
and it has the following useful expression in terms of the spectral traces defined in (\ref{spec-tr}): 
\be
\label{jtrace}
\CJ(\mu, \hbar)=  - \sum_{\ell=1}^\infty Z_\ell {(-\kappa) ^\ell \over \ell}.
\ee
There are certain special combinations of the traces which appear when one expands the spectral determinant around $\kappa=0$: 
\be
\label{xiz}
\Xi(\kappa, \hbar)= 1+ \sum_{N\ge1} Z(N, \hbar) \kappa^N.  
\ee
We will call the $Z(N, \hbar)$, for $N=1,2, \cdots$, the (canonical) partition functions 
associated to the operator $\hat \rho$. We can obtain $Z(N, \hbar)$ by taking an appropriate residue at the origin, 
\be
\label{zn-res}
Z(N, \hbar)= \int_{-\pi \ri}^{\pi \ri}{\rd \mu \over 2 \pi \ri} \re^{\CJ (\mu,\hbar) - N \mu}. 
\ee
If we denote by 
\be
\rho(x_1, x_2) = \langle x_1 |\hat \rho |x_2 \rangle, 
\ee
then the $Z(N, \hbar)$ can be interpreted as the canonical partition functions of an ideal Fermi gas of $N$ particles with energy levels $E_n$:
\be
\label{znmm}
Z(N, \hbar)= {1 \over N!} \sum_{\sigma  \in S_N} (-1)^{\epsilon(\sigma)}  \int  \rd ^N x \prod_i \rho(x_i, x_{\sigma(i)}). 
\ee
In this equation, $S_N$ is the permutation group of $N$ elements and $\epsilon (\sigma)$ is the signature of a permutation $\sigma \in S_N$. The canonical 
partition functions encode the information in the spectral traces in a slightly different way, as one can see by combining (\ref{xiz}) with (\ref{jtrace}), and they are related by  
\be
\label{conjclasses}
Z(N, \hbar) =\sum_{\{ m_\ell \}} {}^{'}\prod_\ell   {(-1)^{(\ell-1)m_\ell } Z_\ell^{m_\ell} \over m_\ell! \ell^{m_\ell}},
\ee
where the $ {}^{'}$ means that the sum is over the integers $m_\ell$ satisfying the constraint
\be
\label{Ncons}
\sum_\ell \ell m_\ell=N.
\ee
We note that the grand potential $\CJ(\mu, \hbar)$ has a well-defined classical limit: when $\hbar \rightarrow 0$, one has 
\be
\label{class-j}
\CJ(\mu, \hbar)={1\over \hbar} \CJ_0(\mu)+\hbar \CJ_1(\mu)+\cdots,
\ee
where the leading contribution
\be
\CJ_0(\mu)=-\sum_{\ell\geq 1}{\frac{(-\kappa)^\ell}{\ell}Z_\ell^{(0)}}
\ee
involves the classical limit of the spectral traces (\ref{cl-traces}). As first noted in \cite{mp}, the study of this limit for the operators appearing 
in Chern--Simons--matter theories leads to many insights on their behavior, see for example \cite{moriyama,mpufu}.

We will now make a proposal for the spectral determinant of the operators $\hat \rho_X$ that we associated to toric CY manifolds. 
We will focus on the case in which the mirror curve has genus one, i.e. on the case of toric (almost) del Pezzo. We sketch 
the generalization to higher genus in the final section of the paper. For simplicity, we will first write down our formulae in the case in which the parameters $m_i$ appearing 
in the operator take their most symmetric value. This value is obtained as follows: the parameters $m_i$ are linear sigma model parameters, 
and they are related to their corresponding K\"ahler parameters or flat coordinates $t_{m_i}$ by an {\it algebraic} mirror map. The most symmetric value of the $m_i$ corresponds to 
setting $t_{m_i}=0$. For example, in the case of local $\IP^1 \times \IP^1$ and local $\IF_1$, the most symmetric value is $m=1$. We will consider the more general case in section \ref{gen-mi}. 

Once we restrict ourselves to the value $t_{m_i}=0$ for the parameters $m_i$, the del Pezzo surfaces considered in the previous section 
have a single modulus $z$, which is related to the modulus $\tilde u$ introduced before as
\be
\label{z-mod}
z={1\over \tilde u^r}. 
\ee
Here, the value of $r$ is determined by the geometry of $X$ (in particular, by the canonical class of $S$.) 
For example, for local $\IP^2$ we have $r=3$, while for local $\IP^1 \times \IP^1$ we have $r=2$
(see Table~\ref{table-ops} for other cases). 
For each of these geometries, there is also a quantum mirror map \cite{acdkv} relating 
the modulus $z$ to a flat coordinate $t$, and of the form 
\be
\label{qmmap}
-t=\log(z)+ \sum_{m\ge 1} \widehat a_m(\hbar) z^m. 
\ee
We will now introduce, in analogy with ABJM theory \cite{hmo3}, an ``effective" $\mu$ parameter 
\be
\label{mueff}
\mu_{\rm eff}= \mu+ {1\over C(\hbar)} J_a (\mu, \hbar), 
\ee
where $J_a(\mu, \hbar)$ is defined by a series expansion 
\be
\label{ja}
J_a(\mu,\hbar)= \sum_{\ell \ge 1} a_\ell (\hbar) \re^{-r \ell \mu }, 
\ee
and $C(\hbar)$ has the form
\be
\label{chbar}
C(\hbar)={C\over 2 \pi \hbar}. 
\ee
The coefficient $C$ is given as follows. Let us consider the volume of the region $\CR(E)$ defined in (\ref{reg-E}), which we will denote as ${\rm vol}_0(E)$. At large $E$, the region becomes polygonal, and 
its volume will behave as 
\be
\label{as-vol}
{\rm vol}_0(E) \approx  C E^2+ 2 \pi \left(B_0-{\pi^2\over 6} C \right) + \CO\left(\re^{-E} \right) \cdots, \qquad E\gg 1. 
\ee
The coefficient $C$ in (\ref{chbar}) is the same one determining the asymptotics of the volume (\ref{as-vol}). It can be easily computed from the polygonal limit of the region $\CR(E)$. 

\begin{example} Let us consider again the case of local $\IP^2$. At large $E$, the region $\CR(E)$ becomes the triangle whose boundaries are appropriate segments of the lines 
\be
x=E, \qquad  p=E, \qquad x+p+E=0, 
\ee
which are read immediately from the tropical limit of the mirror curve. The area of this triangle is $9E^2/2$, so we conclude that
\be
C(\hbar)={9 \over 4 \pi \hbar}. 
\ee
We will verify this value with other techniques later on. \qed
\end{example}

The coefficients $a_m (\hbar)$ appearing in (\ref{ja}) are determined by the quantum mirror map (\ref{qmmap}), as follows
\be
\label{alqmm}
a_\ell (\hbar) =-{C(\hbar) \over r} \widehat a_\ell (\hbar). 
\ee
Note from (\ref{qmmap}) and (\ref{ja}) that the complex modulus (\ref{z-mod}) is identified with 
\be
z=\re^{-r \mu}. 
\ee
This is natural, since the chemical potential $\mu$ plays the r\^ole of the energy, and the above relation follows from (\ref{z-mod}) and (\ref{ure}).

 We are now ready to introduce the crucial quantity determining the spectral determinant. In analogy with \cite{hmo2,hmmo,cgm}, 
 we will call it the {\it modified grand potential}. It is essentially the 
 non-perturbative topological string free energy introduced in \cite{hmmo}, and it has the structure
\be
\label{gpmueff}
J_X(\mu, \hbar)=J^{(\rm p)}(\mu_{\rm eff},\hbar) + J_{\rm M2}(\mu_{\rm eff},\hbar)
+J_{\rm WS} (\mu_{\rm eff},\hbar),
\ee
In this equation, the perturbative piece is given by
\be
\label{jper}
J^{(\rm p)}(\mu,\hbar)= {C(\hbar) \over 3} \mu^3 + B(\hbar) \mu + A(\hbar). 
\ee
The coefficient $B(\hbar)$ has the structure
\be
B(\hbar)= {B_0 \over \hbar} + B_1 \hbar, 
\ee
where $B_0$ is the coefficient appearing in the sub-leading asymptotics of ${\rm vol}_0(E)$, in (\ref{as-vol}). The coefficient $B_1$ can be determined from the 
first quantum correction to the B-period, as in the calculations of \cite{mp,km,hw}. The coefficient $A(\hbar)$ is more difficult to determine, although in some special cases 
it can be guessed and/or computed numerically. It can be also fixed by a normalization condition, as we will see in a moment. 
However, since it is independent of $\mu$, it plays a relatively minor r\^ole. In particular, it does not enter into the quantization condition. The function $J_{\rm M2}(\mu_{\rm eff}, \hbar)$ 
has the structure
\be
J_{\rm M2}(\mu_{\rm eff}, \hbar)= \mu_{\rm eff} \widetilde{J}_b(\mu_{\rm eff},\hbar)+\widetilde{J}_c(\mu_{\rm eff},\hbar), 
\ee
where $\widetilde{J}_b$ and $\widetilde{J}_c$ are given by, 
\be
\label{jb-jc}
\ba 
 \widetilde{J}_b(\mu_{\rm eff},\hbar)= \sum_{\ell \ge 1} \tilde b_\ell (\hbar) \re^{-r \ell \mu_{\rm eff} },  \\
  \widetilde{J}_c(\mu_{\rm eff},\hbar)= \sum_{\ell \ge 1} \tilde c_\ell (\hbar) \re^{-r \ell \mu_{\rm eff} }. 
 \ea
 \ee
The coefficients $\tilde b_\ell(\hbar)$ are determined by the so-called refined BPS invariants of $X$ \cite{gv,ikv,ckk}, which we will denote by $N_{j_L, j_R}^d$. Here, 
$d$ is a positive integer which denotes the degree w.r.t. the flat coordinate or K\"ahler modulus $t$ in (\ref{qmmap}), 
and $j_L$, $j_R$ are the spins of the corresponding BPS multiplets. We have the following expression, 
\be
\label{blj}
\widetilde{b}_\ell(\hbar)=-\frac{r \ell}{4 \pi}\sum_{j_L,j_R}\sum_{\ell=dw}\sum_{d}N^{d}_{j_L,j_R}
\frac{\sin\frac{\hbar w}{2}(2j_L+1)\sin\frac{\hbar w}{2}(2j_R+1)}{w^2\sin^3\frac{\hbar w}{2}}. 
\ee
Note that our conventions for the $N^d_{j_L, j_R}$ are as in \cite{hmmo} (in particular, they do not include the sign $(-1)^{2 j_L + 2 j_R}$.)
The coefficients $\tilde c_\ell (\hbar)$ are determined by a generalization of the relationship found in \cite{hmo3} for ABJM theory, 
\be
\label{bcrel}
\tilde c_\ell (\hbar)= -{\hbar^2  \over r \ell } {\partial \over \partial \hbar} \left( {\tilde b_\ell (\hbar) \over \hbar} \right). 
\ee
Finally, the worldsheet instanton part of the modified grand potential is defined by 
\be
\label{jws}
J_{\rm WS}(\mu, \hbar)= \sum_{m \ge 1} d_m (\hbar) (-1)^{ B m} \re^{-2 \pi m r \mu/\hbar}, 
\ee
where $d_m(\hbar)$ is also determined by the BPS invariants, 
\be
 \label{dmk}
d_m (\hbar)=\sum_{j_L,j_R}\sum_{m=d v} \sum_d N^{d}_{j_L,j_R} 
\frac{2j_R+1}{v \left(2\sin\frac{2\pi^2 v}{\hbar}\right)^2}\frac{\sin\left( \frac{4\pi^2 v}{\hbar}(2j_L+1) \right)}{\sin\frac{4\pi^2 v}{\hbar}}, 
\ee
and the $B$-field $B$ in (\ref{jws}) is such that 
\be
\label{bfield}
(-1)^{2j_L + 2j_R-1}= (-1)^{B d} 
\ee
for all the values of $d$, $j_L$, $j_R$ which lead to a non zero BPS invariant $N^d_{j_L, j_R}$. There is a geometric argument, 
explained in \cite{hmmo}, which shows that there is a natural choice of $B$ field which guarantees (\ref{bfield}). In the toric del Pezzo's that we are considering, we 
can set $B=r$, since they are both determined by the canonical class of $S$. It is important to notice that the combinations of BPS invariants which 
enter into the modified grand potential are very specific. Namely, the combination entering 
in (\ref{jws}) involves only the Gopakumar--Vafa invariants $n_g^d$ appearing in the standard topological string \cite{gv}, 
\be
\label{dm-gv}
d_m(\hbar)=\sum_{g\ge 0} \sum_{m=d v} \sum_d n^d_g {1\over v} \left( 2 \sin {2 \pi^2 v\over \hbar} \right)^{2g-2}, 
\ee
while (\ref{blj}) involves the combination of the invariants appearing in the NS limit of the refined topological string. Indeed, in this limit, the instanton 
part of the topological string free energy can be written as\footnote{This differs 
from the convention used in \cite{hmmo} in a factor of $\ri$.}
\be
\label{NS-j}
F^{\rm inst}_{\rm NS}(t, \hbar) =\sum_{j_L, j_R} \sum_{\ell=wd } 
N^{d}_{j_L, j_R}  \frac{\sin\frac{\hbar w}{2}(2j_L+1)\sin\frac{\hbar w}{2}(2j_R+1)}{2 w^2 \sin^3\frac{\hbar w}{2}} \re^{-\ell t}, 
\ee
and we conclude that
\be
\label{FNS-der}
 \widetilde{J}_b(\mu_{\rm eff}, \hbar)= {r \over 2 \pi} \partial_t F^{\rm inst}_{\rm NS}(t, \hbar) \bigg|_{t=r \mu_{\rm eff}}, 
 \ee
i.e. $ \widetilde{J}_b$ is essentially the quantum B-period of \cite{acdkv}. 

One of the most important aspects of the grand potential (\ref{gpmueff}) is the following: the worldsheet instanton piece 
$J_{\rm WS} (\mu_{\rm eff}, \hbar)$ has double poles when $\hbar$ is of the form $2 \pi$ 
times a rational number. The functions $\widetilde{J}_b$ and $\widetilde{J}_c$ have poles at the same values. 
However, in the total function $J_X(\mu, \hbar)$ these poles {\it cancel}. The proof of 
this statement is a trivial generalization of the proof offerered in \cite{hmmo}, but we present it here for the convenience of the reader, since it is an important point 
of the construction. The coefficient \eqref{dmk} has double poles when $\hbar \in 2 \pi v/\IN$. The coefficient (\ref{blj}) has 
a simple pole when $\hbar  \in 2 \pi \IN/w$, and due to (\ref{bcrel}) 
the coefficient $\tilde c_\ell (\hbar)$ will have a double pole at the same values of $\hbar$. These poles contribute to terms of the same order in $\re^{-\mu_{\rm eff}}$
precisely when $\hbar$ takes the form
\be
\label{ksing}
\hbar={2 \pi v \over w}={2 \pi m \over \ell}.
\ee
 We have then to examine the pole structure of (\ref{gpmueff}) at these values of $\hbar$. Since both (\ref{dmk}) and (\ref{blj}) involve a sum over BPS multiplets with quantum numbers 
 $d$, $(j_L, j_R)$, we can look at the contribution to the pole structure of each multiplet. In the worldsheet instanton contribution, the singular part associated to a BPS multiplet around $\hbar =2v \pi/w$ is 
 given by 
 \be
 \label{ws-pole}
{(-1)^{Bm} \over \pi}  \left[  {v \pi \over  w^4 \left( \hbar  -{2 \pi v \over w}\right)^2 } 
+ {1\over    \hbar  -{2 \pi v \over w}}  \left( {1\over w^3} +   {m r   \mu_{\rm eff}  \over  2 v w^2}   \right) \right] (1+2 j_L)(1+ 2j_R) 
N^{d}_{j_L,j_R} \re^{-{ m r w\over v}  \mu_{\rm eff} }.
 \ee
The singular part in $\mu_{\rm eff} \widetilde J_b (\mu_{\rm eff},k) $ associated to a BPS multiplet is given by 
\be
\label{b-pole}
-{  1 \over 2 \pi}  {\ell r \over w^3 \left( \hbar  -{2 \pi v \over w} \right)} (-1)^{v(2j_L+2j_R-1)} (1+2 j_L)(1+ 2j_R)  N^{d}_{j_L,j_R}\mu_{\rm eff}\re^{-r \ell\mu_{\rm eff}}. 
\ee
Using (\ref{bcrel}), we find that the corresponding singular part in $\widetilde J_c (\mu_{\rm eff},\hbar) $ is given by 
\be
\label{c-pole}
-{1 \over \pi} \left[  {v \pi \over  w^4 \left( \hbar-{2 \pi v \over w}\right)^2 } 
+ {1\over    w^3 \left( \hbar -{2 \pi v \over w}  \right)} \right] (-1)^{v(2j_L+2j_R-1)} (1+2 j_L)(1+ 2j_R)  N^{d}_{j_L,j_R}\re^{-r \ell\mu_{\rm eff}}.
\ee
By using (\ref{bfield}), it is easy to see that all poles in (\ref{ws-pole}) cancel against the poles in (\ref{b-pole}) and (\ref{c-pole}), 
for any value of $\mu_{\rm eff}$. This cancellation 
phenomenon was of course one of the guiding principles for the proposal of \cite{hmmo} and it generalizes the 
HMO cancellation mechanism for the modified grand potential of ABJM theory \cite{hmo2}.

\begin{figure}
\center
\includegraphics[height=6cm]{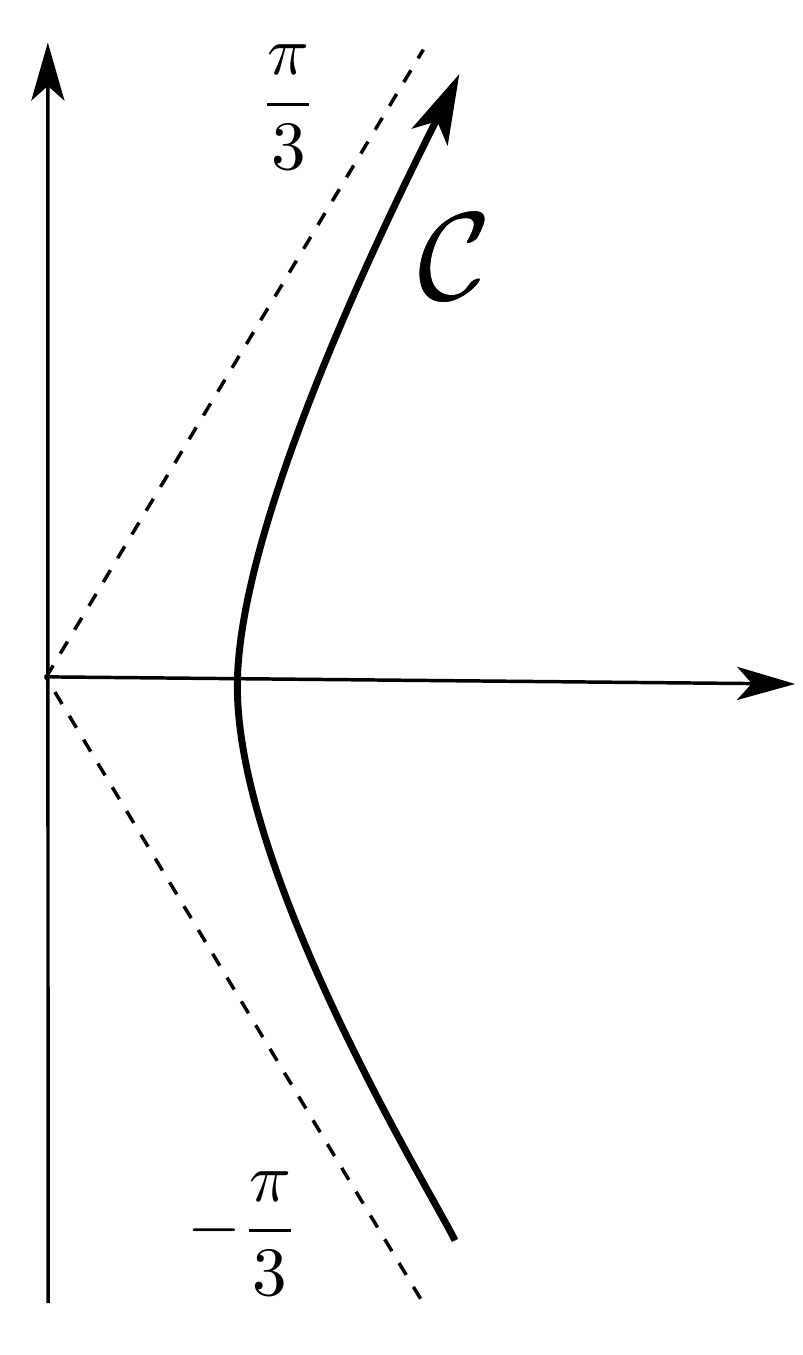}  
\caption{The contour $\CC$ in the complex plane of the chemical potential, which can be used to calculate the canonical partition function from the modified grand potential.}
\label{airy-c}
\end{figure}

We are now ready to make our main proposal: we conjecture that, given a toric del Pezzo CY $X$, the spectral determinant of the operator $\hat \rho_X$ associated to it is given by 
\be
\label{spec-conj}
\Xi_X(\mu, \hbar)= \re^{J_X(\mu, \hbar)} \Theta_X(\mu, \hbar),
\ee
where $J_X(\mu, \hbar)$ is the modified grand potential (\ref{gpmueff}), and $\Theta_X(\mu, \hbar)$ is given by 
\be
\label{ggtf}
\ba
\Theta_X(\mu, \hbar)= \sum_{n \in \IZ} \exp \biggl\{ & - 4\pi^2 n^2 C(\hbar) \mu_{\rm eff} + 2 \pi \ri n ( C(\hbar) \mu_{\rm eff}^2+ B(\hbar)) - {8 \pi^3 \ri n^3 \over 3} C(\hbar) \\
& + 2 \pi \ri n \widetilde{J}_b (\mu_{\rm eff}, \hbar) + J_{\rm WS}(\mu_{\rm eff} + 2 \pi \ri n, \hbar) - J_{\rm WS}(\mu_{\rm eff}, \hbar)\biggr\}. 
\ea
\ee
We will refer to this quantity as the {\it generalized theta function} associated to $X$. The reason for this name is that, in some special cases, it actually becomes 
a theta function, as we will see. Notice that we can write 
\be
\label{xi-j}
\Xi_X(\mu, \hbar)= \sum_{n \in \IZ} \re^{J_X(\mu+2 \pi \ri n, \hbar)}, 
\ee
and it leads to a periodic function of $\mu$. This type of relationship between the grand canonical partition function and 
the modified grand potential was proposed in \cite{hmo2} and recently 
exploited in \cite{cgm} to obtain many new results in $\CN=8$ ABJ(M) theories. It also leads to a very useful formula for the canonical 
partition function as a contour integral: we can use (\ref{xi-j}) to replace $\CJ_X(\mu, \hbar)$ by the modified grand potential in the integrand of (\ref{zn-res}), 
and to extend simultaneously the integration contour along the full imaginary axis. We then deform it to the contour $\CC$ shown in 
\figref{airy-c}, which is the appropriate one in view of the cubic behavior in $\mu$ of $J_X(\mu, \hbar)$, and is in fact the contour used to define the 
Airy function, as in \cite{mp}. We finally obtain the contour integral representation, 
\be
\label{zn-airy}
Z_X(N, \hbar)={1\over 2 \pi \ri} \int_\CC \re^{J_X(\mu,\hbar) - N \mu} \rd \mu. 
\ee
Another way to obtain the canonical partition functions is by simply expanding the spectral determinant 
around $\kappa=0$. This corresponds to $\tilde u\rightarrow 0$, which is the 
point
\be
z=\infty 
\ee
in the moduli space of the CY. This is usually the orbifold limit of the geometry. Therefore, 
according to our conjecture, the spectral traces of the operator $\hat \rho_X$ are determined by topological string theory near the 
orbifold point. We will see some concrete examples of how this works in the examples. 

We would like to note that our proposal can be already tested at the semiclassical level. Indeed, it is easy to see that the WKB expansion of $\CJ_X(\mu, \hbar)=\log \Xi_X(\mu, \hbar)$ is given, according to our 
conjecture, by 
\be
\CJ^{\rm WKB}_X(\mu, \hbar)= J^{({\rm p})}(\mu_{\rm eff}, \hbar) +J_{\rm M2}(\mu_{\rm eff}, \hbar). 
\ee
The l.h.s. of this equation can be in principle computed systematically as in (\ref{class-j}), and this should be reproduced by the expansion of the r.h.s. around $\hbar=0$. We will see examples of 
this later on. 

Let us make some clarifications on the analytic properties of the functions that we have introduced. First of all, note that we have defined the modified grand potential based on an expansion at 
large $\mu$, which corresponds to the large radius expansion of topological string theory. Since this function involves the all-genus free energy of the topological string, one could 
suspect that it leads to a divergent expansion. However, extensive evidence based on concrete examples shows that, when $\hbar$ is real, the modified grand potential $J_X(\mu, \hbar)$ 
is analytic around $\mu=\infty$ \cite{hmmo,cgm}, i.e. it is analytic in a region 
of the form 
\be
{\rm Re}(\mu) > \mu_*.
\ee
Similarly, the generalized theta function (\ref{ggtf}) seems to be analytic in the same region. On the other hand, and as we have 
mentioned above, the spectral determinant of a trace class operator is an entire function on the 
fugacity plane. Therefore, if our conjecture (\ref{spec-conj}) is true, the product in (\ref{spec-conj}), which involves two functions which 
are analytic only in a region of the fugacity plane, is entire\footnote{The fact that the product of a theta function with an appropriate factor leads to an entire function is not unheard of. 
It happens for example in the analysis of blowup functions in Donaldson--Witten theory, see for example \cite{mmwhitham} for a review and references.}. Finally, the 
canonical partition function $Z_X(N, \hbar)$ is only defined in principle for positive integer $N$. However, 
by using the Airy type of integral in (\ref{zn-airy}), we can extend it to an entire function on the complex 
plane of the $N$ variable, exactly as argued in \cite{cgm}. Note that, in this formalism, the value 
of $Z_X(0, \hbar)$ is naturally fixed to be one
\be
\label{normz}
Z_X(0, \hbar)=1, 
\ee
since this is the first term in the expansion of the spectral determinant (\ref{xiz}). This can be used as a normalization condition which fixes completely the $\mu$-independent function $A(\hbar)$.

\subsection{The quantization condition} 

The first piece of information that we can extract from the spectral determinant (\ref{spec-conj}) 
is the spectrum of the operator, which can be read from its zeros. 
Let us then analyze the zeros of (\ref{spec-conj}). This function is the product of two factors: the first factor 
behaves as $\exp(\mu^3)$, while the second one, which we have called a generalized theta function, is oscillating. Therefore, it is 
natural to search for the spectrum by looking at the zeros of this generalized theta function. To search for the zeros, we write, as suggested by 
(\ref{kappa-E}), 
\be
\mu=E + \pi \ri, 
\ee
therefore 
\be
\mu_{\rm eff}= E_{\rm eff}+ \pi \ri, 
\ee
where
\be
\label{Eeff}
E_{\rm eff}= E +{1\over C(\hbar)} J_a (E+ \pi \ri). 
\ee
Note that this introduces a sign depending on the parity of $rm$, 
\be
\label{Jasign}
J_a (E+ \pi \ri)=\sum_{m \ge 1} (-1)^{rm } a_m (\hbar) \re^{-r m E}. 
\ee
We then find
\be
\label{theta12}
\ba
\Theta_X(E+\pi \ri, \hbar)= \re^{\zeta} \sum_{n \in \IZ} \exp \biggl\{  &- 4\pi^2 (n+1/2)^2 C(\hbar) E_{\rm eff} 
 - {8 \pi^3 \ri (n+1/2)^3 \over 3} C(\hbar) \\
 &+ 2 \pi \ri (n +1/2)\left( C(\hbar) E_{\rm eff}^2+ B(\hbar) +\tilde J_b (E_{\rm eff}+ \pi \ri, \hbar) \right)\\
&+ f_{\rm WS}(E_{\rm eff}+ \pi \ri, n)-{1\over 2} f_{\rm WS} (E_{\rm eff}+ \pi \ri,-1)\ \biggr\}. 
\ea
\ee
In this equation, we have introduced the functions, 
 \be
 \ba
 f_{\rm WS}(\mu, n)&= J_{\rm WS}(\mu+ 2\pi \ri n, \hbar)- J_{\rm WS}(\mu, \hbar)\\
 &=\sum_{m \ge 1} d_m (\hbar)\left( \re^{-4 \pi^2 \ri m r n /\hbar}-1\right) (-1)^{B m} \re^{-2 \pi m r \mu/\hbar}, 
 \ea
 \ee
 for $n\not=0$, and by definition $f_{\rm WS}(\mu, 0)=0$. We have, in particular,  
 \be
f_{\rm WS} (E_{\rm eff}+ \pi \ri,-1)= 2 \ri  \sum_{m \ge 1} d_m (\hbar)\sin {2 \pi^2 m r \over \hbar}  (-1)^{B m} \re^{-2 \pi m r E_{\rm eff} /\hbar}. 
\ee
The overall factor $\zeta$ is given by 
\be
\ba
\zeta&= \pi^2 C(\hbar) E_{\rm eff} -\pi \ri \left( C(\hbar) E_{\rm eff}^2+ B(\hbar) +\tilde J_b (E_{\rm eff}+ \pi \ri, \hbar)\right) \\
&+{1\over 2} f_{\rm WS}(E_{\rm eff}+ \pi \ri,-1) + {\pi^3 \ri C(k) \over 3}. 
\ea
\ee
%
To extract a quantization condition from this equation, we will think of the generalized theta function as a sum of exponentially small corrections, in which the leading 
order is given by the terms $n=0,-1$ in (\ref{theta12}). If we keep only these two terms, we see that (\ref{theta12}) is given by 
\be
\label{cosO}
\exp \left(\zeta - \pi^2 C(\hbar) E_{\rm eff}^2 \right)\cos \left( \pi \Omega(E) \right)
\ee
where 
\be
\label{approx}
\Omega(E)= \Omega_{\rm p}(E)+ \Omega_{\rm np}(E), 
\ee
and 
\be
\label{pnp}
\ba
\Omega_{\rm p}(E)&= C(\hbar) E_{\rm eff}^2 + B(\hbar)-{\pi^2 \over 3} C(\hbar) + \tilde J_b (E_{\rm eff}+ \pi \ri), \\
\Omega_{\rm np}(E)&= -{1\over \pi} \sum_{m \ge 1} d_m (\hbar)\sin {2 \pi^2 m r \over \hbar}  (-1)^{B m} \re^{-2 \pi m r E_{\rm eff} /\hbar}. 
\ea
\ee
In this approximation, in which we keep only the first two terms in the generalized theta function, the quantization condition 
reads
\be
\label{approxWKB}
\Omega(E)=  s+{1\over 2}, \qquad s=0,1,2, \cdots
\ee
Although (\ref{cosO}) also vanishes for negative, integer $s$, the condition (\ref{approxWKB}) does not seem to have solutions in $E$ for those values.

Let us pause a moment to examine this quantization condition. It has a perturbative part in 
$\hbar$, given by $\Omega_{\rm p}(E)$, and a non-perturbative part given by $\Omega_{\rm np}(E)$. The perturbative part is what one would find 
by using just the NS limit of the refined topological string, or the perturbative WKB approach of \cite{acdkv}. As pointed out in \cite{km}, this perturbative quantization 
condition {\it can not} be the whole story: the operator $\hat \rho_X$ has a well-defined spectrum at values of $\hbar$ of the form $2 \pi$ 
times a rational number, but for these values of $\hbar$ the perturbative part has 
poles. Therefore, the perturbative approach is fundamentally incomplete. 
As pointed out in \cite{km}, one should include instanton corrections, and these should cancel the poles in the perturbative part. The proposal of \cite{km} 
for these non-perturbative corrections is in fact to add $\Omega_{\rm np}(E)$ to the perturbative function $\Omega_{\rm p}(E)$, as in (\ref{approx}), 
so that the modified quantization condition is (\ref{approxWKB}). This condition was originally 
proposed for ABJM theory, which is a particular case of the above construction, and then extended to ABJ theory in \cite{kallen}. 

The quantization condition (\ref{approxWKB}) of \cite{km} has two virtues: first of all, in contrast to the 
perturbative WKB condition, it makes sense for any real value of $\hbar$. Second, it 
reproduces the spectrum of the operators in some special cases. However, it doesn't lead to the 
right energies for generic values of $\hbar$. This was noted experimentally in some examples in \cite{hw}, 
based on an extensive numerical analysis (see also \cite{fhw}). But it is now clear why this is so: the quantization condition 
(\ref{approxWKB}) has corrections due to higher order terms in the generalized theta function (\ref{theta12}). These 
corrections can be determined analytically, as follows. Let us write the exact quantization condition as 
\be
\label{exact-qc}
\Omega(E)+ \lambda(E)= s+{1\over 2}, \qquad s=0,1,2, \cdots, 
\ee
where $\lambda(E)$ is the sought-for correction. Let us denote 
\be
\label{fs-fc}
  \ba
  f_{\rm c}(n)&=\sum_{m \ge 1} (-1)^{B m} d_m (\hbar)\left( \cos \left( {2 \pi^2 r m (2n+1) \over \hbar} \right) -\cos \left({2 \pi^2 r m  \over \hbar}  \right)\right) \re^{-2\pi r m E_{\rm eff} /\hbar}, \\
    f_{\rm s}(n)&=\sum_{m \ge 1} (-1)^{B m} d_m (\hbar)\left( \sin \left({2  \pi^2 r m (2n+1) \over \hbar} \right)  -(2n+1) \sin \left( {2 \pi^2 r m  \over \hbar} \right) \right) \re^{-2\pi r m E_{\rm eff} /\hbar}, 
    \ea
    \ee
 for $n\not=0$, and $f_{\rm c}(0)=f_{\rm s}(0)=0$. Note that, as functions of $\hbar$, they do not have singularities, and they are 
 determined by the Gopakumar--Vafa invariants entering into $d_m(\hbar)$. A simple calculation shows that $\lambda(E)$ is determined by the equation 
    \be
    \label{qc-corr}
    \ba
&  \sum_{n =0}^\infty \re^{-4 \pi^2 n(n+1) C(\hbar) E_{\rm eff} } (-1)^n\re^{f_{\rm c}(n)} \\
  & \qquad\qquad \times  \sin \left({4 \pi^3 n(n+1)(2n+1) \over 3} C(\hbar)+  f_{\rm s}(n) + 2 \pi (n+1/2) \lambda(E) \right)=0.  
   \ea
   \ee
Although this equation looks complicated, $\lambda(E)$ can be 
obtained as a power series in the small parameter 
\be
\exp(- 2 \pi C E_{\rm eff}/\hbar). 
\ee
We have written $C(\hbar)$ as in (\ref{chbar}) to make manifest that all these corrections are non-perturbative in $\hbar$. 
The zeroth order approximation is obtained by picking just $n=0$ in the above sum, which leads to $\lambda(E)=0$, 
and we reproduce (\ref{approxWKB}). The leading non-trivial correction is  
\be
\lambda(E)\approx {1\over \pi}\exp(- 4 \pi C E_{\rm eff}/\hbar) \re^{f_{\rm c}(1)}  \sin \left(8 \pi^3 C(\hbar)+  f_{\rm s}(1) \right), 
\ee
which has itself an expansion in powers of $\exp\left( -2 \pi r E_{\rm eff}/\hbar\right)$ due to the factors of $f_{\rm c}(1)$, $f_{\rm s}(1)$. 
As we will see in concrete examples, this reproduces the proposed corrections in \cite{hw}, and therefore it agrees with the numerical results obtained so far 
for the spectrum of the operators.

\subsection{Physical interpretation} 

Let us now pause a little bit to comment on the physical significance of the above conjectures for a description of the topological string. 

First of all, let us understand in detail how the standard perturbative expansion of the topological string emerges from this picture. The modified grand potential 
$J_X(\mu, \hbar)$ can be studied in various regimes. 
In the semiclassical regime, we have $\mu$ fixed and $\hbar \rightarrow 0$. But as pointed out in \cite{kkn,mp,gm}, 
there is a 't Hooft limit given by  
\be
\label{muthooft}
\hbar \rightarrow \infty, \qquad \hat \mu= {\mu \over \hbar} \quad {\text{fixed}}. 
\ee
In this regime, the modified grand potential has an expansion of the form
\be
\label{jthooft}
J_X^{\text {'t Hooft}}(\mu, \hbar)= \sum_{g\ge 0} \hbar^{2-2g} J_X^{(g)}\left( \hat \mu \right), 
\ee
which selects precisely the worldsheet instanton part $J_{\rm WS}(\mu, \hbar)$ (in this limit, $\mu_{\rm eff}= \mu$.)
 Indeed, if we assume that the function $A(\hbar)$ has an asymptotic expansion in this regime of the form 
\be
A(\hbar)= \sum_{g\ge 0} A_g \hbar^{2-2g}, 
\ee
we find that the $J_X^{(g)}(\hat \mu)$ are essentially the genus $g$ free energies of the standard topological string at large radius. We have, for example, 
\be
\ba
J_X^{(0)} \left( \hat \mu \right)&= {C\over 6 \pi } \hat \mu^3 + B_1 \hat \mu +A_0+ {1\over 16 \pi^4} \sum_{w,d\ge 1} n_0^d {(-1)^{Bd w} \over w^3} \re^{- 2 \pi r w d \hat \mu}, \\
J_X^{(1)} \left( \hat \mu \right)&=B_0 \hat \mu +  A_1+\sum_{w,d\ge 1} \left( {n_0^d  \over 12} + n_1^d\right) {(-1)^{Bd w} \over w} \re^{- 2 \pi r w d \hat \mu}. 
\ea
\ee
The basic quantity in our approach is the spectral determinant $\Xi_{X}(\mu, \hbar)$. It follows from our main conjecture that, in the 't Hooft limit (\ref{muthooft}), the spectral determinant has the 
asymptotic expansion
\be
\label{thooft-asym}
\log \Xi_{X}(\mu, \hbar) \sim J_X^{\text {'t Hooft}}(\mu, \hbar). 
\ee
The theta function gives an oscillatory, non-perturbative correction which is similar to the oscillatory corrections to the large $N$ asymptotics of matrix models \cite{bde}. 

In terms of the canonical partition function $Z_X(N, \hbar)$, the regime (\ref{muthooft}) corresponds to the standard 't Hooft regime
\be
\label{Nthooft}
\hbar \rightarrow \infty, \qquad  {N \over \hbar} \quad {\text{fixed}}. 
\ee
In this regime, $Z_X(N, \hbar)$ has an expansion at strong 't Hooft coupling which is obtained by a Laplace 
or Fourier transform of (\ref{jthooft}), as it follows from (\ref{zn-airy}). The expansion (\ref{jthooft}) indicates as well that the parameter 
$\hbar$ plays the r\^ole of the {\it inverse} topological string coupling constant, 
\be
\label{gtop}
g_{\rm top} \sim {1\over \hbar}. 
\ee

The above results are structurally very similar to what has been obtained for 
ABJM theory, and our conjectures have been inspired by the structure of the ABJM partition function on $\IS^3$. 
Indeed, what we are proposing is an interpretation of the topological string as an ideal Fermi gas, where the Hamiltonian $\hat H_X$ is given by (\ref{hamiltonian}). This Fermi gas provides a 
microscopic description of the topological string, which is weakly coupled when the topological string is strongly coupled, due to (\ref{gtop}) (that such a description 
should exist was already anticipated in \cite{mm-talk}, based on the results of \cite{mp}.) The perturbative 
genus expansion emerges as a particular asymptotic expansion of this microscopic description, as we have seen above.  

This Fermi gas picture of the topological string has various important properties, which we now comment in some detail. 

 First of all, it includes {\it non-perturbative effects} in the topological string coupling constant. These effects are encoded in 
the functions $J_a$, $\widetilde J_b$ and $\widetilde J_c$, which come from the refined topological string in the NS limit. 
This of course was already pointed out in \cite{hmmo}. Conversely, from the dual point of view of the spectral problem, it is the worldsheet instanton 
contribution which leads to non-perturbative effects in $\hbar$, as explained in \cite{km}. 

Second, our description is {\it M-theoretic}, 
in the same way that the Fermi gas approach to ABJM theory captures its M-theory regime. In particular, our description involves in a crucial way an M-theoretic aspect of topological 
string theory, which is the Gopakumar--Vafa resummation of the genus expansion. We need this resummation in order to find results at finite $\hbar$. At the same time, 
in our picture, this resummation 
is not enough, and in particular it can not be used to analyze the spectral problem, due to the 
presence of poles. To cancel these poles we need, as in the HMO mechanism \cite{hmo2}, the non-perturbative contributions encoded in the 
NS limit of the refined string. 

Third, our description is {\it background independent}, in the following sense. The perturbative topological string free energy depends on a choice of 
``duality frame," and different frames are appropriate for different regions of moduli space. This is reflected in the fact that the genus $g$ free energies have a relatively complicated 
analytic structure, displaying branch cuts which lead to different ``phases" \cite{witten-phases}. In contrast, in our Fermi gas approach, the basic object 
is the spectral determinant or grand canonical partition function. Since the operators we are considering seem to be of trace class, the spectral determinant 
is an entire function on the fugacity plane. From the point of view of the topological string, this means that it is an {\it entire function on the CY moduli space} parametrized 
by $\tilde u$. In particular, it does not depend on the choice of frame (although its asymptotic expansions in 
different regimes might pick a convenient frame, like in (\ref{thooft-asym})). Of course, the modified grand potential is not an entire function: it rather has a complicated analytic structure, inherited from the non-trivial analytic structure of the CY periods. 
However, if our conjecture is true, the inclusion of the generalized theta function (\ref{ggtf}) as in (\ref{spec-conj}) leads 
to an entire function. This is of course reminiscent of the proposal of \cite{em} for 
a background independent partition function for topological strings on local CYs. In that paper, and based on previous results \cite{bde,eynard}, it was noticed that including a theta function with a similar 
structure than (\ref{ggtf}) led to a function which was essentially modular invariant\footnote{The fact that background independent formulations of topological string theory should involve 
theta functions in some way or another goes back of course to \cite{witten-bi}. See for example \cite{gnp,dvvonk} for related discussions.}. However, the 
``non-perturbative partition function" constructed in \cite{em} by including the theta function is only defined as a formal expansion in $1/N$, while the r.h.s of (\ref{spec-conj}) is 
well-defined in a region of the $\mu$ plane and for any real value of $\hbar$, and it should extend to an entire function. 

Finally, and on a more speculative note, our description suggests that the underlying microscopic theory behind the operator $\hat \rho_X$ is a theory of $N$ M2 branes, which 
provides a holographic description of topological strings. A first piece of evidence for this speculation is that the canonical free energy, defined by 
\be
\label{cfe}
F_X(N, \hbar) =- \log Z_X(N, \hbar)
\ee
has a universal large $N$ behavior of the form, 
\be
F_X(N, \hbar) \approx {2\over 3} {\sqrt{ {2 \pi \over C}}} \hbar^{1/2} N^{3/2},  \qquad N \gg 1,
\ee
where $C$ is the constant defined by (\ref{as-vol}). This can be deduced from (\ref{zn-airy}) by using the techniques of \cite{mp}. 
Of course, this is the expected 
behavior in a theory of $N$ M2 branes \cite{kt}. In relation to this, recall that, if $Y_8$ is a cone over a Sasaki--Einstein manifold $X_7$, 
and we put $N$ M2 branes on 
\be
\label{y8}
\IR^3 \times Y_8, 
\ee
located at the tip of the cone, this background is described at large $N$ by M-theory on 
\be
\label{ads4}
{\text {AdS}}_4 \times X_7. 
\ee
In \cite{dvv} it was suggested that topological string theory on the CY $X$ is defined by M-theory on the background 
\be
\label{dvv-back}
TN \times (X \times \IS^1), 
\ee
where $TN$ is the four-dimensional Taub--Nut space. It might happen that the background (\ref{dvv-back}) emerges by backreaction of $N$ M2 branes in a related space, in the same 
way as (\ref{ads4}) emerges from (\ref{y8}). If the proposal of \cite{dvv} is correct, this approach might give a hint of what is this theory of M2 branes.

\subsection{The maximally supersymmetric cases}

There are some special values of $\hbar$ for which the general results presented above simplify considerably. The modified grand potential becomes simpler when 
\be
\label{susy-values}
\hbar =\pi \quad \text{or} \quad  \hbar=2 \pi. 
\ee
We will refer to these two cases as the ``maximally supersymmetric cases," in analogy with what happens 
in ABJM theory, where these values correspond to the enhancement of supersymmetry from $\CN=6$ to $\CN=8$. This is 
the situation analyzed in \cite{cgm}, and the analysis of this subsection is very similar to what was done in that paper. 
The reason why the supersymmetric cases are special is that, in those cases, all the contributions to $d_m(\hbar)$ in (\ref{dm-gv}) with $g \ge 2$ vanish, and we only have 
contributions of the conventional topological string up to one-loop. Similarly, the contributions coming from the refined topological string involve the $\hbar$ expansion of the 
NS limit up to next-to-leading order. Finally, the generalized theta function (\ref{ggtf}) becomes a standard Jacobi theta function. 

We will now present some general formulae for the grand potential and the spectral determinant in the case $\hbar=2 \pi$. The case with $\hbar=\pi$ is similar and can be 
worked out as in \cite{cgm}. 

Let us first analyze the behavior of the coefficients 
$\tilde b_\ell(\hbar)$ and $\tilde c_\ell(\hbar)$ as $\hbar \rightarrow 2 \pi$. They will have a singular part, and a finite part. The singular part 
will cancel against similar contributions in the worldsheet instantons, by the generalized HMO mechanism. It is easy to see that the coefficient $\widetilde{b}_\ell(\hbar)$ has the following behavior 
as $\hbar \rightarrow 2 \pi$:
\be
\widetilde{b}_\ell(\hbar)= {\widetilde{b}^{-1} _\ell \over \xi} + \widetilde{b}^{1}_\ell \xi+\CO(\xi^2) , \qquad \xi= \hbar-2 \pi,
\ee
therefore its finite part vanishes when $\hbar \rightarrow 2 \pi$. The behavior of $\widetilde c_\ell (\hbar)$ is, from (\ref{bcrel}), 
\be
\widetilde{c}_\ell(\hbar)={\widetilde{c}^{-2} _\ell \over \xi^2}+{\widetilde{c}^{-1} _\ell \over \xi}-{2 \pi \over r \ell} \widetilde{b}^{1}_\ell+\CO(\xi).
\ee
From (\ref{blj}) we find the following expression,  
\be
\label{b-limit}
 \widetilde{b}^{1}_\ell=\frac{r \ell}{48 \pi }\sum_{j_L,j_R}\sum_{\ell=dw}N^{d}_{j_L,j_R} {(-1)^{B\ell } \over w} m_L m_R \left( -3+ m_L^2 + m_R^2 \right), 
 \ee
 where we have denoted
 \be
 m_L=2j_L+1, \qquad m_R = 2j_R+1, 
 \ee
 and we have taken into account the relationship (\ref{bfield}). We would like to express the above quantity in terms of functions known in closed form. 
To do this, we compare the BPS expansion of the refined free energy in the NS limit, given in (\ref{NS-j}), 
to its perturbative expansion in $\hbar$. We also have to take into account the term $(-1)^{B\ell}$ in (\ref{b-limit}). If we write, 
\be
\label{ns-expansion}
F^{\rm inst}_{\rm NS}(t+\pi \ri B, \hbar)=\sum_{n \ge 0} \hbar^{2n-1} \widehat F_n^{\rm NS, inst} (t),
\ee
we deduce that the finite part of $\widetilde{J}_c(\mu_{\rm eff})$ as $\hbar \rightarrow 2 \pi$ is simply
\be
\label{jc-ns}
\widehat F^{\text{NS, inst}}_{1} (t).
\ee
As in (\ref{FNS-der}), we have to identify $t=r\mu_{\rm eff}$. Note that this free energy differs from the usual one in the shift of $t$ by a $B$ field, as in (\ref{jws}). 

Let us now consider the worldhseet instanton part. As mentioned before, all the terms in $d_m (\hbar)$ with $g\ge 2$ vanish. 
The $g=1$ contribution survives in the limit $\hbar \rightarrow 2 \pi$, 
and we have to keep the finite part of $g=0$. A simple calculation shows that the finite part as $\hbar \rightarrow2 \pi$ of 
\be
 \left( 2 \sin {2 \pi^2 w \over \hbar } \right)^{-2} {\rm e}^{-{2  \pi r d w \mu \over \hbar}}
\ee
is 
\be
{{\rm e}^{-r d w \mu} \over 12 \pi^2 w^2} \left( 3+ \pi^2 w^2 + 3 r dw \mu + {3 \over 2} d^2 w^2 r^2 \mu^2\right). 
\ee
The finite piece of (\ref{jws}) as $\hbar \rightarrow 2 \pi$ is then, 
\be
 {r^2 \mu_{\rm eff}^2 \over 8 \pi^2} \partial^2_t \widehat F^{\rm inst}_0(t) - {r \mu_{\rm eff} \over 4 \pi^2} \partial_t  \widehat F^{\rm inst}_0(t) +{1\over 4 \pi^2} \widehat F^{\rm inst}_0(t) + 
 \widehat F^{\rm inst}_1(t), 
\ee
where we denoted
\be
\widehat F^{\rm inst}_0(t)= \sum_{w,d \ge1} n^d_0  {(-1)^{wd B} \over w^3} \re^{-d w t}, 
\qquad \widehat F^{\rm inst}_1(t)= \sum_{w,d \ge1}  \left( {n^d_0 \over 12} + n^d_1\right) {(-1)^{wd B} \over w}  \re^{-d w t}. 
\ee
These are the genus zero and genus one free energies of the standard topological string, but with the inclusion of an extra $B$-field. 
Here, and for the moment being, we only keep the instanton part of these free energies (i.e. we drop all the 
polynomial parts in $t$), and we use that $t=r \mu_{\rm eff}$. We conclude that, 
\be
\label{full-formula}
\ba
J_X(\mu, \hbar=2 \pi)&=  J^{(\rm p)}(\mu_{\rm eff},2 \pi)+{r^2 \mu_{\rm eff}^2 \over 8 \pi^2} \partial^2_t \widehat F^{\rm inst}_0(t) - {r \mu_{\rm eff} \over 4 \pi^2} \partial_t  \widehat F^{\rm inst}_0(t) +{1\over 4 \pi^2} \widehat F^{\rm inst}_0(t) \\
&+ \widehat F^{\rm inst}_1(t)+
  \widehat F^{\text {NS, inst}}_1 (t).
\ea
\ee
A more compact expression is obtained if we introduce the full prepotential, 
\be
\widehat F_0(t)= {C \over 3 r^3} t^3+ \widehat F_0^{\rm inst}(t). 
\ee
Then, we can write
\be
\label{mgp-2}
\ba
J_X(\mu, \hbar=2 \pi)&=A(2 \pi) +{1\over 4 \pi^2} \left(\widehat F_0(t) - t \partial_t \widehat F_0(t) + {t^2 \over 2} \partial_t^2  \widehat F_0(t)\right)\\
&  + {B(2 \pi) \over r} t +\widehat  F^{\rm inst}_1(t)+
\widehat  F^{\text {NS, inst}}_1 (t),
  \ea 
  \ee
 where we have taken into account (\ref{jper}) and (\ref{chbar}). Like before, we have to set $t=r \mu_{\rm eff}$. 
 
 It is also easy to obtain the generalized theta function in the case $\hbar=2 \pi$. One finds, 
 \be
 \label{tt}
 \Theta_X(\mu, 2 \pi)= \sum_{n \in \IZ} \exp \biggl\{  {\pi \ri n^2 r^2\over 4}  \tau + 2 \pi \ri n  \left( \xi + B(2 \pi) \right) - {2  \pi \ri n^3 C \over 3 } \biggr\},
 \ee
 where
 \be
 \label{tau-f}
 \tau={2 \ri \over \pi} \partial_t^2 \widehat F_0(t)
 \ee
 and 
 \be
 \label{xi-f}
 \xi= {r\over 4 \pi^2} \left( t \partial_t^2  \widehat F_0(t) -\partial_t  \widehat F_0(t)\right). 
 \ee
 Although (\ref{tt}) does not look like a theta function, it can be reduced to one if $C$ is an integer or half-integer. Indeed, since 
 \be
 {n(n^2-1) \over 3} 
 \ee
 is even for any $n \in \IZ$, we can write (\ref{tt}) as 
 \be
  \Theta_X(\mu, 2 \pi)= \sum_{n \in \IZ} \exp \biggl\{  {\pi \ri n^2 r^2\over 4}  \tau + 2 \pi \ri n  \left( \xi + B(2 \pi) - {C \over 3} \right)  \biggr\},
  \ee
  which is a standard Jacobi theta function,
  \be
    \Theta_X(\mu, 2 \pi)= \vartheta_3\left( v, {r^2 \tau \over 4}\right), 
    \ee
    with 
    \be
    v= \xi + B(2 \pi) - {C \over 3}. 
    \ee

  Note that, due to the properties of special geometry, we have that ${\rm Im}(\tau)>0$, therefore the above 
  theta function is well defined. The spectral determinant is given by 
  \be
  \label{ms-xi}
  \Xi_X(\mu, 2 \pi)=\re^{J_X(\mu, 2 \pi)} \vartheta_3\left( v, {r^2 \tau \over 4}\right). 
  \ee
  This is similar to the result obtained in \cite{cgm} for maximally supersymmetric ABJ(M) theories. It was shown in \cite{em} that the combination 
  \be
  \label{em-f}
  \exp\left( {1\over 4 \pi^2} \left(\widehat F_0(t) - t \partial_t\widehat F_0(t) + {t^2 \over 2} \partial_t^2\widehat F_0(t) \right) +\widehat  F_1(t) \right) \vartheta\left[^\alpha_\beta\right]\left(v, {r^2 \tau \over 4}\right), 
  \ee
 involving a general theta function with characteristics, is essentially invariant under modular transformations. The 
 exponent in (\ref{em-f}), involving $J_X(\mu,  2\pi)$, is slightly different from the one in (\ref{em-f}). However, in all examples we have studied, this difference is 
 a modular invariant function of $z$, therefore (\ref{ms-xi}) inherits the modular properties of (\ref{em-f}). 
 We conclude that, in the maximally supersymmetric case, our conjectural expression for the spectral determinant is 
  given by a modular invariant expression. We now have a natural explanation for this property: it is due to the fact that the spectral determinant is an entire 
  function on the complex moduli space of the CY $X$. 
  
  Finally, let us consider the quantization condition in the maximally supersymmetric cases. 
  It is easy to see that, in those cases, the function $f_s(n)$ defined in (\ref{fs-fc}) vanishes for all $n=1,2,\cdots$. In addition, 
  if $C$ is a half-integer, the first term in the argument of the sine in the second line of (\ref{qc-corr}) is always an integer multiple of $\pi$. Therefore, the solution to 
  (\ref{qc-corr}) is $\lambda(E)=0$ and there are no corrections 
  to the quantization condition (\ref{approxWKB}) of \cite{km}.  As in \cite{cgm}, we can now write the quantization condition for $\hbar=2 \pi$ in terms of the prepotential. A simple calculation from (\ref{pnp}) gives
  \be
  \label{qc-max}
  CE_{\rm eff}^2 + 4 \pi^2 B(2 \pi)- {\pi^2 C\over 3} + r^2 E_{\rm eff} \partial_t^2 F_0^{\rm inst}(t) - r \partial_t F_0^{\rm inst}(t)= 4 \pi^2 \left( s+{1\over 2} \right), \quad  s=0,1,2, \cdots
  \ee
  where as usual we set $t=r E_{\rm eff}$. 
  
  It is easy to see that there can be other values of $\hbar$ for which the corrections (\ref{qc-corr}) vanish. For example, if $\hbar= \pi s$, 
  where $s$ is a divisor of $2r$, $f_{\rm s}(n)$ is also zero. Although the vanishing of $\lambda(E)$ also depends on the value of $C$, it can be seen that, in all 
  examples, one has again $\lambda(E)=0$ for these values of $\hbar$. However, the modified grand potential will still have higher genus corrections.

\subsection{The case with general parameters}
\label{gen-mi}
So far we have restricted ourselves to the case in which the values of the parameters $m_i$ are such that their corresponding K\"ahler parameters $t_{m_i}$ vanish. 
The general case is a straightforward generalization of the above results. We will denote 
\be
Q_{m_i}= \re^{-t_{m_i}}. 
\ee
Let us also denote the K\"ahler parameters of $X$ by $t_i$, in an arbitrary basis (the choice of basis can be dictated for example by the 
geometry of the CY.) They can be always written down as linear combinations of the K\"ahler parameter which corresponds to $\tilde u$, and the $t_{m_i}$. 
The K\"ahler parameter associated to $\tilde u$ (which is the true modulus of the geometry) should be set to $\mu_{\rm eff}$, and 
we will write 
\be
t_i = c_i \mu_{\rm eff}- \alpha_{ij} \log Q_{m_j}, 
\ee
where $c_i$, $\alpha_{ij}$ depend on the geometry. For example, for local $\IP^1 \times \IP^1$, we have one single parameter 
$Q_m=m$, and 
\be
t_1 = 2 \mu_{\rm eff} - \log m, \qquad t_2 = 2 \mu_{\rm eff}. 
\ee
The appropriate generalization of our conjecture for the modified grand potential is already implicit in the proposal of \cite{hmmo}. 
Let us consider the NS limit of the topological string free energy, which we write as
\be
\label{NS-j-m}
F^{\rm inst}_{\rm NS}({\bf t}, \hbar) =\sum_{j_L, j_R} \sum_{w, {\bf d} } 
N^{{\bf d}}_{j_L, j_R}  \frac{\sin\frac{\hbar w}{2}(2j_L+1)\sin\frac{\hbar w}{2}(2j_R+1)}{2 w^2 \sin^3\frac{\hbar w}{2}} \re^{-w {\bf d}\cdot{\bf  t}}, 
\ee
where ${\bf t}$ is the vector of K\"ahler parameters, and ${\bf d}$ is the vector of degrees. We now introduce a variable $\lambda_s$ and consider the function 
\be
 \label{NS-j-v}
F^{\rm inst}_{\rm NS}({\bf T}, \lambda_s) =\sum_{j_L, j_R} \sum_{w, {\bf d} } 
N^{{\bf d}}_{j_L, j_R}  \frac{\sin\frac{\pi  w}{\lambda_s }(2j_L+1)\sin\frac{\pi w}{\lambda_s }(2j_R+1)}{2 w^2 \sin^3\frac{\pi w}{\lambda_s}} \re^{-w {\bf d}\cdot{\bf  T}/\lambda_s }. 
\ee
Note that this is equivalent to introduce
\be
\label{Tis}
T_i= {2 \pi \over \hbar} t_i
\ee
and set 
\be
\lambda_s={2 \pi \over \hbar}. 
\ee
Then, let us define 
\be
J_{\rm M2}(\mu, \hbar)= -{1\over 2 \pi} {\partial \over \partial \lambda_s} \left( \lambda_s F^{\rm inst}_{\rm NS}({\bf T}, \lambda_s) \right)\biggl|_{\lambda_s={2 \pi \over \hbar}}.
\ee
In taking the derivative, we assume that  $T_i$ are independent of $\lambda_s$. One finds
 \be
 J_{\rm M2}(\mu_{\rm eff}, m_i, \hbar)= \mu_{\rm eff}    \widetilde{J}_b(\mu_{\rm eff},m_i, \hbar)+\widetilde{J}_c(\mu_{\rm eff},m_i, \hbar), 
 \ee
where
 \be
 \ba
 \widetilde{J}_b(\mu_{\rm eff},m_i, \hbar)&=-{1\over 2 \pi} \sum_{j_L, j_R} \sum_{w, {\bf d} }\left(  {\bf c} \cdot {\bf d} \right) N^{{\bf d}}_{j_L, j_R} 
 \frac{\sin\frac{\hbar w}{2}(2j_L+1)\sin\frac{\hbar w}{2}(2j_R+1)}{2 w \sin^3\frac{\hbar w}{2}} \re^{-w {\bf d}\cdot{\bf  t}}, \\
  \widetilde{J}_c(\mu_{\rm eff},m_i, \hbar)&={1\over 2 \pi} \sum_{i,j} \sum_{j_L, j_R} \sum_{w, {\bf d} } d_i \alpha_{ij} \log Q_{m_j} N^{{\bf d}}_{j_L, j_R} 
 \frac{\sin\frac{\hbar w}{2}(2j_L+1)\sin\frac{\hbar w}{2}(2j_R+1)}{2 w \sin^3\frac{\hbar w}{2}} \re^{-w {\bf d}\cdot{\bf  t}}\\
 &+{1\over 2 \pi} \sum_{j_L, j_R} \sum_{w, {\bf d} } \hbar^2 {\partial \over \partial \hbar} \left[  \frac{\sin\frac{\hbar w}{2}(2j_L+1)\sin\frac{\hbar w}{2}(2j_R+1)}{2 \hbar w^2 \sin^3\frac{\hbar w}{2}} \right] N^{{\bf d}}_{j_L, j_R} \re^{-w {\bf d}\cdot{\bf  t}}. 
 \ea
 \ee
 The grand potential is now given by 
 \be
 \label{jx-masses}
 J_X (\mu, m_i, \hbar) = J^{({\rm p})} (\mu_{\rm eff}, m_i, \hbar) + J_{\rm M2} (\mu_{\rm eff}, m_i, \hbar) + J_{\rm WS}(\mu_{\rm eff}, m_i, \hbar), 
 \ee
 where $J^{({\rm p})} (\mu_{\rm eff}, m_i, \hbar) $ is the perturbative part of the grand potential, which might involve now quadratic terms in $\mu^2$, 
 \be
 J^{({\rm p})} (\mu, m_i, \hbar)={C(\hbar) \over 3} \mu^3 + D(m_i, \hbar) \mu^2 + B(m_i, \hbar) \mu + A(m_i, \hbar). 
 \ee
 This corresponds to the fact that the perturbative genus zero and genus one free energies are in general cubic 
 and linear polynomials in the $t_i$, respectively. In (\ref{jx-masses}), 
\be
J_{\rm WS}(\mu_{\rm eff}, m_i, \hbar)= \sum_{g\ge 0} \sum_{{\bf d}, v} n^{\bf d}_g {1\over v} \left( 2 \sin {2 \pi^2 v\over \hbar} \right)^{2g-2} \re^{- v {\bf d} \cdot \left({ 2 \pi  \over \hbar}{\bf t} + \pi \ri {\bf K}\right)}.
\ee
Here, ${\bf K}$ is the vector representing the canonical class of $X$, in the homology basis chosen to represent the $t_i$. As explained in \cite{hmmo}, this 
is needed to implement the cancellation of poles. 

 The above formulae generalize the results presented before for the general case in which $t_{m_i} \not=0$. The spectral determinant is defined again by (\ref{xi-j}), and 
 it is straightforward to write quantization conditions and explicit formulae in the maximally supersymmetric cases from (\ref{jx-masses}). A more detailed study of this general case will appear 
 elsewhere. 

\sectiono{The case of local $\IP^2$}

In the previous section we have presented our conjecture in some generality. We will now perform a detailed analysis of the benchmark example 
for any statement about local mirror symmetry, namely local $\IP^2$. We will first get some 
intuition and useful data from a semiclassical analysis\footnote{Some of these results were obtained 
in the fall of 2013 in \cite{gkm}.}. We will 
then derive the quantization condition for the spectrum in the general case, and we will recover analytically all the results obtained in \cite{hw} by numerical fitting. Finally, we will 
focus on the maximally supersymmetric case and give direct evidence that the spectral determinant is indeed given by (\ref{spec-conj}).

\subsection{Semiclassical analysis} 
Let us then study the operator (\ref{lp2-op}). A very important source of information on this operator is obtained from 
its semiclassical limit, which can be analyzed as in \cite{mp}. In particular, we would like to compute the classical limit of the 
grand potential, given in (\ref{class-j}). It turns out that, in this 
case, the semiclassical traces (\ref{cl-traces}) can be computed in closed form, 
\be
\label{cltr-p2}
Z_\ell^{(0)}=\frac{\Gamma(\frac{\ell}{3})^3}{6\pi \Gamma(\ell)},
\ee
and one finds the explicit formula
\be
\ba
\CJ_0(\mu)&={\kappa   \over 36 \pi}  \biggl\{ 6 \Gamma \left(\frac{1}{3}\right)^3 \, _3F_2\left(\frac{1}{3},\frac{1}{3},\frac{1}{3};\frac{2}{3},\frac{4}{3};-\frac{\kappa ^3}{27}\right)\\
& \quad +\kappa  \left(\kappa  \, _4F_3\left(1,1,1,1;\frac{4}{3},\frac{5}{3},2;-\frac{\kappa ^3}{27}\right)-3 \Gamma \left(\frac{2}{3}\right)^3 \,
   _3F_2\left(\frac{2}{3},\frac{2}{3},\frac{2}{3};\frac{4}{3},\frac{5}{3};-\frac{\kappa ^3}{27}\right)\right)\biggr\}. 
   \ea
   \label{eq:J0-exact}
   \ee
  Expanding around $\kappa=\infty$ we obtain
\be
\CJ_0(\mu)=\frac{3}{4\pi }\mu^3+\frac{\pi}{2}\mu+\frac{4 \zeta(3)}{3\pi}+\left(\frac{9}{2\pi}\mu^2-\frac{9}{2\pi}\mu +\pi-\frac{3}{\pi} \right)\re^{-3\mu}+\mathcal{O}(\re^{-6\mu})~.
\ee
As derived in appendix~\ref{sec:J1}, the first correction $\CJ_1(\mu)$ in \eqref{class-j} is given by
\be
\CJ_1(\mu)=-\frac{1}{72} \mathcal{J}_0''(\mu).
\label{eq:J1}
\ee
Thus one finds
\be
\mathcal{J}_1(\mu)=-\frac{\mu}{16\pi}+\( -\frac{9}{16\pi}\mu^2+\frac{21}{16\pi}\mu -\frac{1}{8\pi} -\frac{\pi}{8} \)\re^{-3\mu}+\cO(\re^{-6\mu}).
\ee
From these formulae we can immediately deduce that
\be
\label{cb-p2}
C(\hbar)= {9 \over 4 \pi \hbar}, \qquad B(\hbar) ={\pi \over 2 \hbar}-{\hbar \over 16 \pi}, 
\ee
and
\be
\label{pah}
A(\hbar)=\frac{4 \zeta(3)}{3\pi \hbar}+\cO(\hbar^3).
\ee
The semiclassical result for the grand potential also makes it possible to verify that (\ref{bcrel}) holds in the limit $\hbar \rightarrow 0$. In addition, it is a testing ground for 
the results for $J(\mu, \hbar)$ at finite $\hbar$, which we now explain. 

\subsection{The grand potential and the quantization condition}

Let us now write down the results for the modified grand potential at finite $\hbar$. Since this will be needed in the following, 
we recall some basic facts about mirror symmetry of local $\IP^2$. In this case, $r=3$, therefore the parameter $z$ is related to $\tilde u$ by 
\be
z=\tilde u^{-3}, 
\ee
and we can identify 
\be
z=\re^{-3 \mu}. 
\ee
We can take then the B-field to be $B=1$. The two basic periods at large radius are given by, 
\be
\ba
\widetilde \varpi_1(z)&= \sum_{j\ge 1} 3 {(3j-1)! \over (j!)^3}  (-z)^j, \\
\widetilde \varpi_2(z)&=\sum_{j \ge 1}{ 18\over j!} 
{  \Gamma( 3j ) 
\over \Gamma (1 + j)^2} \left\{ \psi(3j) - \psi (j+1)\right\}(-z)^{j}. 
\ea
\ee
The prepotential is defined by the standard relations, 
\be
\ba
Q&=\re^{-t}= z \exp \left( \widetilde \varpi_1(z) \right)=z- 6 z^2+\cdots, \\ 
\partial_t F_0(t)&= {1\over 6} \left( \log^2(z) + 2  \widetilde \varpi_1(z) \log(z) +  \widetilde \varpi_2(z) \right), 
\ea
\ee
which leads to 
\be
\label{fp2}
F_0(t)={t^3 \over 18} +F^{\rm inst}_0(t), \qquad F_0^{\rm inst}(t)=3 Q- {45 \over 8} Q^2+{244 \over 9} Q^3-\cdots.
\ee
Note that, when computing the modified grand potential, $t$ is given by $r\mu_{\rm eff}$ or $rE_{\rm eff}$, which depends explicitly on $\hbar$. 

Let us now write down the modified grand potential. The first ingredient we need is $\mu_{\rm eff}$. By using the quantum mirror map of local $\IP^2$ \cite{acdkv}, 
the relation (\ref{alqmm}), and (\ref{mueff}), we find that
\be
\mu_{\rm eff}= \mu+ {4 \pi \hbar \over 9} J_a(\mu)=  \mu+ (q^{1/2}+ q^{-1/2}) \re^{-3 \mu} -  \left( 6 + {7 \over 2}(  q+ q^{-1}) + q^{2}+ q^{-2} \right)\re^{-6 \mu}+ \cdots
\ee
where $q$ is defined by (\ref{q-def}). One can check that the limit $\hbar \rightarrow 0$ of this expression reproduces the result for the 
$a_\ell$ coefficients of the semiclassical grand potential.The coefficients $C(\hbar)$, $B(\hbar)$ in the 
perturbative part (\ref{jper}) can be read from (\ref{cb-p2}). 
 The series appearing in (\ref{jb-jc}) can be also computed explicitly from (\ref{blj}) and (\ref{bcrel}), and 
they read, to the very first orders, 
\be
\ba
\widetilde J_b (\mu_{\rm eff}, \hbar)&= -{3 \over 4 \pi}   (2 \cos (\hbar )+1) \csc \left(\frac{\hbar
   }{2}\right)  \re^{-3 \mu_{\rm eff}}\\
   &  + \frac{3}{8 \pi } \sin (3 \hbar ) \left(4 \csc ^2\left(\frac{\hbar }{2}\right)-\csc ^2(\hbar )\right)
   \re^{-6 \mu_{\rm eff}}+\cdots, \\
 \widetilde J_c (\mu_{\rm eff}, \hbar)&=  \frac{1}{8 \pi } \csc ^2\left(\frac{\hbar }{2}\right) \left(-2 \sin \left(\frac{3 \hbar }{2}\right)-4 \hbar  \cos \left(\frac{\hbar }{2}\right)+\hbar  \cos
   \left(\frac{3 \hbar }{2}\right)\right)  \re^{-3 \mu_{\rm eff}}+\cdots .
   \ea
   \ee
Finally, the worldsheet instanton part of the modified grand potential 
is determined by the Gopakumar--Vafa invariants of local $\IP^2$, which are given by 
\be
n_0^1=3, \quad n_0^2=-6, \cdots, 
\ee
and $J_{\rm WS}(\mu, \hbar)$ reads 
\be
J_{\rm WS}(\mu, \hbar)= -3 \left( 2 \sin {2 \pi^2 \over \hbar} \right)^{-2} \re^{-6 \pi \mu/\hbar} +\cdots. 
\ee
The only ingredient in the modified grand potential which we have not specified is $A(\hbar)$, since for the moment being our theory does not determine it its general form. 
However, we have the following educated guess for it. Let 
 \be
\label{ak}
A_{\rm c}(k)= \frac{2\zeta(3)}{\pi^2 k}\left(1-\frac{k^3}{16}\right)
+\frac{k^2}{\pi^2} \int_0^\infty \frac{x}{\re^{k x}-1}\log(1-\re^{-2x})\rd x.
\ee
be the function appearing in the modified grand potential of ABJM theory. This function first appeared in the Fermi gas formulation of \cite{mp}, and 
a closed form expression for it was found in \cite{hanada} by using the constant map contribution to the topological string 
free energy. This form was further simplified in \cite{hatsuda-o} to (\ref{ak}). Then, we propose that the $A(\hbar)$ function of local $\IP^2$ is given by 
\be
\label{ah-p2}
A(\hbar) ={3 A_{\rm c}(\hbar/\pi)- A_{\rm c}(3\hbar/\pi) \over 4}. 
\ee
Using the known results for $A_{\rm c}(k)$, it is easy to see that this function has the small $\hbar$ expansion (\ref{pah}). In addition, we have verified numerically for many values of $\hbar$ that 
it leads to the normalization condition (\ref{normz}). 
%
%

We are now ready to analyze the quantization condition determining the spectrum of the operator (\ref{lp2-op}). 
The exact result is given in (\ref{exact-qc}), where $\Omega(E)$ is given by the approximate 
quantization condition of \cite{km}, and $\lambda(E)$ can be determined from (\ref{qc-corr}) as a 
power series in $\exp(-6 \pi E_{\rm eff}/\hbar)$. We get, 
\be 
\label{qcp2-corr}
\lambda(E)=\lambda_1 \re^{-18\pi E_{\rm eff}/\hbar}+\lambda_2\re^{-24 \pi E_{\rm eff}/\hbar}+\lambda_3\re^{-30\pi  E_{\rm eff}/\hbar}+\lambda_4\re^{-36\pi  E_{\rm eff}/\hbar}+\mathcal{O}\left(  \re^{-42 \pi E_{\rm eff}  /\hbar}\right) ,
\ee
with 
\be 
\ba
\lambda_1& =\frac{1}{\pi} \sin \left(18 x\right), \\
\lambda_2 &= \frac{3}{\pi} \sin ^2\left(6x \right) \sin \left(24 x \right) \csc ^2\left(2x \right), \\
\lambda_3 &= \frac{3}{\pi} \sin \left(6x \right) \sin \left(30 x\right) \csc ^2\left(2x\right) \\
& \times \left(16 \sin \left(2x \right) 
\sin ^2\left(6x\right)+20 \sin \left(2x\right) \sin \left(10 x\right) \sin \left(6 x \right)+7 \sin \left(18 x\right)\right),
\\
\lambda_4&=  \frac{\csc ^2(2 x)}{8 \pi } \Big( -36 \sin (4 x)-74 \sin (8 x)-140 \sin (12 x)-146 \sin (16 x)-184 \sin (20 x) \\
&\qquad -64 \sin (24 x)+68 \sin (28 x)+391 \sin (32 x)+478 \sin (36 x)+391 \sin (40 x)
\\ & \qquad+68 \sin (44 x)-64 \sin (48 x) -184 \sin (52 x)-146 \sin (56 x)-140 \sin (60 x)\\ 
&\qquad-74 \sin (64 x) -56 \sin (68 x)-14 \sin (72 x)-20 \sin (76 x)\Big),\\
 \ea
 \ee
 and we have denoted
 \be
 \label{xvar}
 x={ \pi^2 \over \hbar}. 
 \ee
 In order to test this corrected quantization condition, we should compute the spectrum of the operator (\ref{lp2-op}) and see if we can reproduce it. 
 Fortunately, this has been done in detail by Huang and Wang in \cite{hw}, where they compute the spectrum numerically for many 
 values of $\hbar$. In their study, they noticed that (\ref{approxWKB}) fails for generic values of $\hbar$, and they computed a series of correction terms by fitting 
 their numerical data. It turns out that the first three terms in (\ref{qcp2-corr}) coincide exactly with the 
 corrections proposed in \cite{hw} from numerical analysis! We conclude that our exact quantization condition reproduces the available numerical data of the spectrum of 
 (\ref{lp2-op}).

 \begin{table}[t] 
\centering
\begin{tabular}{l  l}
\hline
Degree &	$E_0$  \\
\hline
   1&      \underline{3.777}6432527296085597046797\\
 3 &       \underline{3.77770625}05593500784461494\\
   6&      \underline{3.77770625858220}08247337693\\
    8 &    \underline{3.77770625858220699}72270331\\
   10&    \underline{3.77770625858220699868}77030\\
   11 &   \underline{3.7777062585822069986880}502\\
   12 &   \underline{3.7777062585822069986880709}\\
\hline
Numerical value & 3.7777062585822069986880709             
\\
\end{tabular}
\caption{The first energy level for  local $\mathbb{P}^2$ with $\hbar=4\pi$ calculated analytically, from the zeros of the generalized theta function. In the first column the degree $d$ indicates that we are including instantons corrections up to $\re^{-3 d E/2}$ in the generalized theta function (\ref{theta12}). In the last line the numerical value is given. }
 \label{4pitable}
\end{table}
\begin{table}[t] 
\centering
\begin{tabular}{l  l}
\hline
Degree &	$E_0$  \\
\hline
   1&   \underline{4.3514}491328672074939635\\
 3 &    \underline{4.35143749}58036556904823\\
   6&   \underline{4.351437497126020}9085142\\
    8 & \underline{4.351437497126020298}0640\\
   9&   \underline{4.351437497126020298101}1\\    
   10&  \underline{4.3514374971260202981017}\\

\hline
Numerical value &    4.3514374971260202981017        
\\
\end{tabular}
\caption{The first energy level for  local $\mathbb{P}^2$ with $\hbar=5\pi$ calculated analytically. In the first column the degree $d$ indicates that we are including instantons corrections up to $\re^{-6 d E/5}$. In the last line the numerical value is given. }
 \label{5pitable}
\end{table}

\begin{figure} \begin{center}
 {\includegraphics[scale=0.6]{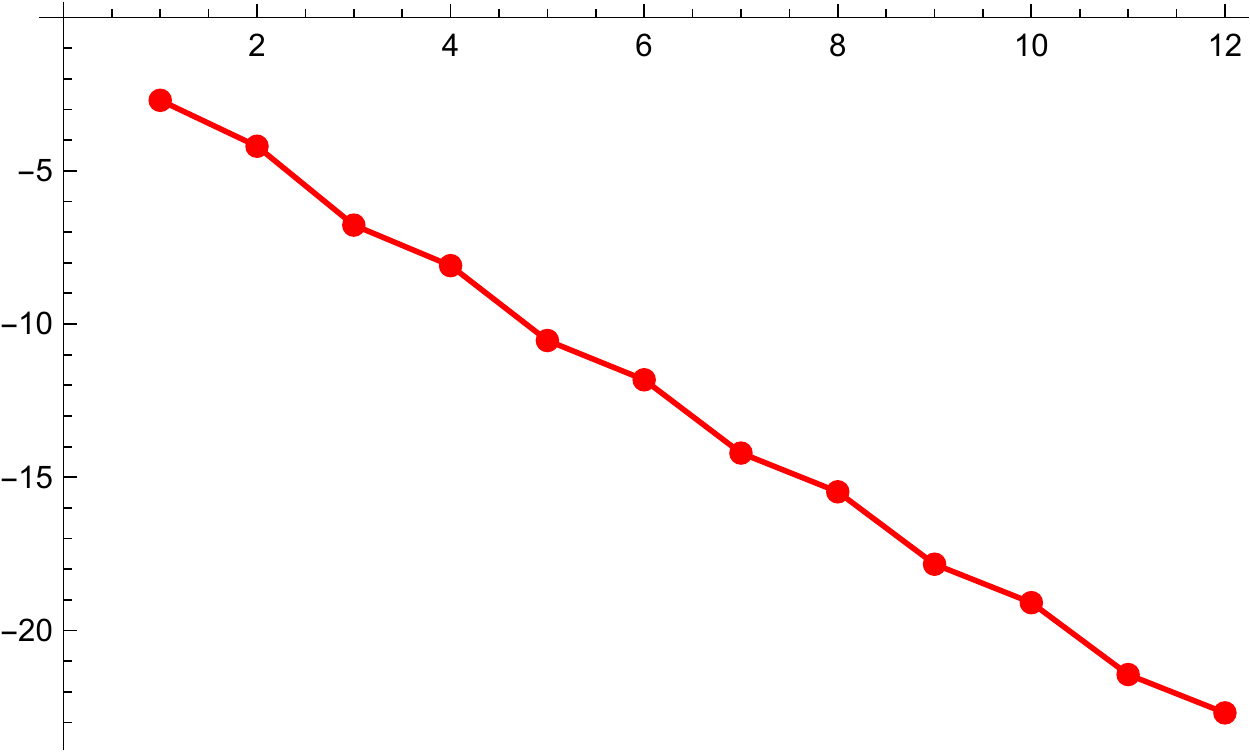} }
\caption{ The difference $\Delta(4\pi,m)$, defined in (\ref{diff}), for the local $\mathbb{P}^2$  geometry.}
 \label{Figp2num}
  \end{center}
\end{figure}

 We have actually improved the numerical analysis of \cite{hw} to further test our conjecture. We first note that the matrix element (\ref{omatrix}) for the operator (\ref{lp2-op}) 
 has the following form:
\be\label{p2m}
M=\begin{pmatrix}
 m_1 & 0 & 0 & m_8& 0 & 0 &m_9 &  \ldots\\
 0 & m_2 & 0 & 0 & m_{10} & 0 & 0 \\
 0 & 0 & m_3 & 0 & 0 & m_{11}  & 0 \\
 m_8 & 0 & 0 & m_4 & 0 & 0 & m_{12}  \\
 0 & m_{10} & 0 & 0 & m_5 & 0 & 0 \\
 0 & 0 &m_{11}  & 0 & 0 & m_6 & 0 \\
m_9& 0 & 0 & m_{12} & 0 & 0 & m_7 \\
\vdots &  & & &  &  &\ddots
\end{pmatrix}~.
\ee
In particular this matrix can be decomposed into three matrices $M^{(0)},M^{(1)}$ and $M^{(2)}$ where
\be M^{(i)}_{mn}=M_{3m+i,3n+i}, \quad i=0,1,2.\ee
Due to the peculiar form (\ref{p2m}), one finds that the eigenvalues of $M^{(i)}$ correspond to 
\be \re^{E_{3n+i}}, \quad n=0,1,\dots \ee
 The $M^{(i)}$  are still infinite-dimensional matrices, however when one truncates them to an $L \times L$ matrix, 
 the corresponding eigenvalues $ E_{3n+i}^{(L)}$ behave as
 \be E_{3n+i}^{(L)}=E_{3n+i}+\mathcal{O}\left({1\over L}\right), \quad L\gg 1, \ee
where $E_{3n+i}$ are the exact eigenvalues. Therefore, one can apply the method of Richardson extrapolation to accelerate the convergence 
of these eigenvalue as $L\rightarrow \infty$ and obtain high precision numerical results.
In Table \ref{4pitable} and \ref{5pitable} we compare the analytic and the numerical results for the ground state energy, for $\hbar=4 \pi$ and $\hbar=5 \pi$, respectively. 
The analytic values are computed by looking at the zeros of the generalized theta function (\ref{theta12}). As expected, the more instantons we include in the analytic computation of the grand potential, 
the better we approach the numerical value. This result can be illustrated by looking at
\be \label{diff} \Delta(\hbar,m)= \log_{10}\left| E^{\rm num}_0(\hbar)-E^{(m)}_0(\hbar) \right|,
\ee
where $E^{\rm num}_0(\hbar)$ is the numerical value of the ground state energy, and $E^{(m)}_0(\hbar)$ is the value computed from (\ref{theta12}) by including the first $m$ instanton corrections. 
As shown in \figref{Figp2num} in the case of $\hbar=4 \pi$, $E^{(m)}_0 (4 \pi)$ converges to $E^{\rm num}_0 (4 \pi)$ as 
$m$ grows. This is precisely what we expect if our conjecture is correct.
%

Let us now consider the quantization condition in the maximally supersymmetric case $\hbar=2\pi$. The quantum mirror map 
simplifies considerably, and we find 
\be
E_{\rm eff}= E- {1\over 3} \widetilde \varpi_1\left(\re^{-3 E} \right). 
\ee
 The quantization condition can be written in terms of the prepotential, as shown in (\ref{qc-max}). After taking into account the various signs, we find that it is given by  
\be
\label{sp-maxp2}
{9\over 2} E_{\rm eff}^2 -\pi^2 - 3 \partial_t  F_0^{\rm inst}(t)+ 9 E_{\rm eff} \partial^2_t  F_0^{\rm inst}(t)=4 \pi^2 \left( s+{1\over 2} \right), \qquad  s=0,1,2,\cdots 
\ee
This expression can be written explicitly in terms of Meijer G-functions.  
From (\ref{sp-maxp2}) it is possible to determine the spectrum with very high accuracy. For example, 
for the ground state energy, and with 25 significant digits, the above quantization condition gives, 
\be
E_0=2.562642068623819370817399...
\ee
which should be compared to the numerical value obtained directly by diagonalizing the matrix (\ref{omatrix}) with size $L=800$, 
\be
E_0^{(800)}=2.5626420686238193708...
\ee
\subsection{The spectral determinant}

We have given what we feel is convincing evidence that our conjecture leads to the right spectrum for the operator (\ref{lp2-op}), 
but our proposal is actually stronger: it leads to an explicit prediction for the 
spectral determinant of (\ref{lp2-op}). We now test this last point in detail. A simple testing ground is the maximally supersymmetric case $\hbar=2 \pi$. 
Our strategy will be the following: first, we will compute the 
spectral traces (\ref{spec-tr}) numerically from the spectrum. Since our quantization condition reproduces the numerical spectrum, we can use (\ref{sp-maxp2}). 
From this and (\ref{conjclasses}) we can compute the canonical partition function $Z(N, \hbar=2\pi)$ for low values of $N$. 
However, according to our conjecture, this can be also computed from the spectral determinant, i.e. from the modified grand potential, by using 
(\ref{xiz}) or (\ref{zn-airy}). Agreement of both calculations gives a strong support to our 
conjecture. This is similar to the procedure followed in \cite{cgm} for the maximally supersymmetric ABJ(M) theories. 

Let us first consider the spectral traces. Although we have computed them numerically, the results can be fitted to exact expressions with high precision. We find, 
\be
\label{p2-traces}
\ba
Z_1&= {1\over 9}, \\
Z_2&={1\over 27}-{1\over 6 \pi{\sqrt{3}} }, \\
Z_3&={1\over 81} - {1\over 24 \pi^2} - {1\over 24\pi{\sqrt{3}}},\\
Z_4&=-{1\over 729} + {1 \over 72 \pi^2}.
\ea
\ee
The fact that the spectral traces have such nice forms already indicates that this is a particularly beautiful spectral theory. 
Let us now see if we can reproduce these results from our 
conjecture (\ref{spec-conj}). We first calculate $J(\mu, 2 \pi)$ from the general formula (\ref{mgp-2}). The effective chemical potential is given by 
\be
\label{mms-p2}
\mu_{\rm eff}= \mu - {1\over 3} \widetilde \varpi_1\left(-\re^{-3 \mu}\right).  
\ee
The genus zero free energy is obtained by mirror symmetry. The genus one free energy has the closed form expression, 
\be
\widehat F_1(t)={1\over 2} \log\left(-{\rd z \over \rd t}  \right)-{1\over 12} \log\left( z^7 \left(1-27 z \right) \right), 
\ee
where we have switched the sign $z\rightarrow -z$ in the standard expressions due to the presence of the $B$ field and of the form 
(\ref{mms-p2}) of the effective chemical potential. The refined  
genus one free energy in the NS limit can be also computed in closed form \cite{hk}, and it is given by 
\be
\widehat F_1^{\rm NS}(t)= -{1\over 24} \log \left({1-27 z \over z}\right), 
\ee
where we have again changed the sign in $z$ due to the non-trivial $B$ field. Notice also that 
\be
\widehat F_1^{\rm NS, inst}(t)=\widehat F_1^{\rm NS}(t)+{t\over 24}, 
\ee
and 
\be
\widehat F_1^{\rm inst}(t)=\widehat  F_1(t) -{t\over 12}. 
\ee
Since 
\be
B(2 \pi)={1\over 8}, 
\ee
we conclude that
\be
\label{j-amxp2}
J(\mu, 2\pi)=  A(2 \pi)+  {1\over 4 \pi^2} \left(\widehat F_0(t) - t \partial_t \widehat F_0(t) + {t^2 \over 2} \partial_t^2 \widehat F_0(t)\right)+\widehat F_1(t)+ \widehat F_1^{\rm NS}(t), 
\ee
and we have to set, 
\be
t=3 \mu_{\rm eff}. 
\ee
The value $A(2 \pi)$ can be found from the conjecture (\ref{ah-p2}) and the results of \cite{hatsuda-o} for the explicit values of $A_{\rm c}(k)$. We find, 
\be
A(2 \pi)= {1\over 6} \log(3)- {\zeta(3) \over 3 \pi^2}, 
\ee
which leads to the normalization condition (\ref{normz}) with high numerical precision. 

With all this information we can already write down the full large $\mu$ expansion of $J(\mu, 2 \pi)$. We find for the very first orders, 
\be
\ba
J(\mu, 2\pi)&=  {3 \mu^3 \over 8 \pi^2} +{\mu \over 8}+ {1\over 6} \log(3)- {\zeta(3) \over 3 \pi^2}+  
\left(-\frac{45 \mu ^2}{8 \pi ^2}-\frac{9 \mu }{4 \pi ^2}-\frac{3}{4 \pi ^2}+\frac{3}{8}\right) \re^{-3 \mu} \\
& + 
 \left(-\frac{999 \mu ^2}{16 \pi ^2}-\frac{63 \mu }{16 \pi ^2}+\frac{9}{32} \left(34-\frac{5}{\pi ^2}\right) \right)\re^{-6 \mu} +\cdots
 \ea
 \ee
 We conclude that the spectral determinant is given by the specialization of (\ref{ms-xi}) to our case, namely
 \be
 \label{spec-lp2}
 \Xi(\kappa, 2 \pi)= \exp\left( J(\mu, 2 \pi) \right) \vartheta_3 \left( \xi-{3 \over 8}, {9 \tau \over 4}\right), \qquad \kappa=\re^\mu.
 \ee

\begin{figure}
\center
\includegraphics[height=5cm]{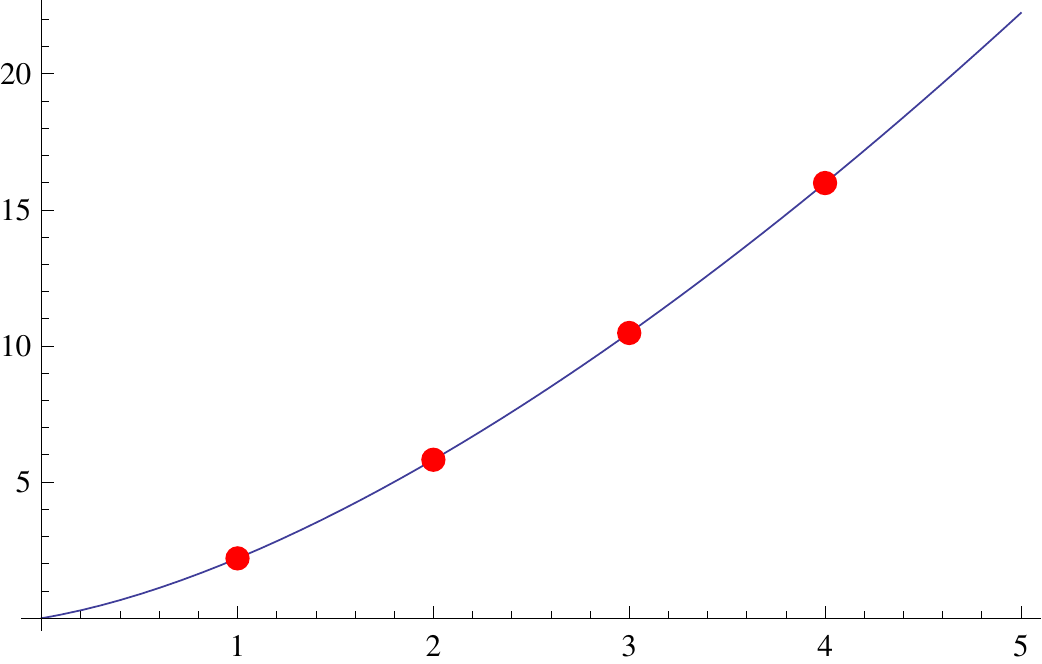}  
\caption{The smooth line gives the free energy $F(N, 2 \pi)$ of local $\IP^2$ as a function of $N$, computed from (\ref{zn-airy}), while the points give the values of 
of the free energy as computed from the spectral traces (\ref{p2-traces}).}
\label{znp2}
\end{figure}
 
We are now ready to verify that (\ref{spec-lp2}) leads to the correct spectral properties. Instead of checking the traces (\ref{p2-traces}), we can 
equivalently check the values of the first few canonical partition functions, which are determined by (\ref{p2-traces}) and (\ref{conjclasses}). We have for example, 
\be
Z(1, 2 \pi)= {1\over 9}, \qquad Z(2, 2 \pi)= \frac{1}{12 \sqrt{3} \pi }-\frac{1}{81}, 
\ee
and so on. These can then be compared to the 
canonical partition functions as computed from (\ref{spec-lp2}). A convenient way to do this computation is, like in \cite{hmo2,cgm}, by using (\ref{zn-airy}). 
The r.h.s. of (\ref{zn-airy}) can be 
computed as a convergent sum of Airy functions and their derivatives. Indeed, let us expand 
\begin{equation}
\re^{J(\mu, 2\pi)} = \re^{J^{({\rm p})} (\mu,  2 \pi)} \sum_{l=0}^{\infty} \re^{-3 l \mu}\sum_{n=0}^{2l} a_{l,n} \mu^n. 
\end{equation}
After integration, the expansion in $\mu$ can translated into derivatives with respect to $N$, and this leads to the expression
\begin{equation}
\label{z-exp}
Z(N,2 \pi)=\frac{\re^{A(2 \pi)}}{\left(C(2 \pi) \right)^{1/3}}
\sum_{l=0}^{\infty} \sum_{n=0}^{2l} a_{l,n} \left(-\frac{\partial}{\partial N}\right)^n \mathrm{Ai}
\left(\frac{N+3 l -B(2\pi)}{\left(C(2\pi)\right)^{1/3}}\right), 
\end{equation}
where ${\rm Ai}(z)$ is the Airy function. This can be computed numerically with a very high precision, 
and we find an impressive agreement with the values of $Z(N, 2 \pi)$ computed directly from the spectrum. 
For example, for $N=1$, the above Airy calculation, including ten terms in the 
expansion in $\re^{-3 \mu}$, agrees with $Z(1, 2 \pi)=1/9$ with a precision of 160 digits. Including more 
corrections increases the precision arbitrarily. In fact, (\ref{zn-airy}) 
gives an interpolating function defined for all complex values of $N$, as in \cite{cgm}. 
In \figref{znp2}, we show the function $-\log Z(N, 2 \pi)$ obtained from (\ref{zn-airy}), as a function of $N$, together with the 
values for $N=1,2,3,4$ obtained from the spectral traces. 

There is an alternative way to obtain the above results for the canonical partition functions, which is simply to expand the spectral determinant (\ref{spec-lp2}) around $\kappa=0$ 
and read the $Z(N, 2\pi)$ from this expansion. This is also geometrically interesting, since it involves local $\IP^2$ near the orbifold point $z=\infty$. In practice, 
to compute this expansion from our formula (\ref{spec-conj}), we must expand separately the functions $J_X(\mu, \hbar)$ and the generalized theta function. 
It turns out that, in this case, both functions have branch cuts in the real $z$ axis, and as we go to the orbifold point we hit the conifold singularity. Therefore, the analytic 
continuation of each factor is not well-defined. However, since the product is itself well-defined, 
this is easily solved: we just have to change $\kappa \rightarrow - \kappa$ in (\ref{spec-lp2}). This only flips the sign in the expansion (\ref{xiz}), which is the expansion of an 
entire function at the origin, but now the analytic continuation from $z=0$ to $z=\infty$ avoids the conifold singularity. 
A simple calculation shows that, after this change, the spectral determinant is given 
by 
\be
\label{xi-orb}
\Xi(-\kappa, 2 \pi)= \exp(J_{\rm orb}(\kappa, 2 \pi)) \Theta_{\rm orb}(\kappa, 2 \pi).  
\ee
In this formula, $J_{\rm orb}(\kappa, 2 \pi)$ is given by the same formula of (\ref{j-amxp2}), but where the hatted free energies are replaced by the conventional free energies. In addition, we have 
\be
t= 3\mu-\widetilde \varpi_1\left(\re^{-3 \mu} \right). 
\ee
The advantage of this formulation is that all the ingredients have now an analytic continuation to the orbifold point $z=\infty$, and one finds (see for 
example \cite{agm,abk}):
\be
\ba
t&=\kappa\frac{  \Gamma \left(\frac{1}{3}\right)
 }{\Gamma
   \left(\frac{2}{3}\right)^2}  \, _3F_2\left(\frac{1}{3},\frac{1}{3},\frac{1}{3};\frac{2}{3},\frac{4}{3}
   ;-\frac{\kappa ^3}{27}\right)-\kappa ^2 \frac{\Gamma
   \left(\frac{2}{3}\right) }{2 \Gamma \left(\frac{1}{3}\right)^2}\,
   _3F_2\left(\frac{2}{3},\frac{2}{3},\frac{2}{3};\frac{4}{3},\frac{5}{3}
   ;-\frac{\kappa ^3}{27}\right),\\
\partial_t F_0(t)&= \kappa \frac{\pi   \Gamma \left(\frac{1}{3}\right) }{3 \sqrt{3} \Gamma
   \left(\frac{2}{3}\right)^2}\, _3F_2\left(\frac{1}{3},\frac{1}{3},\frac{1}{3};\frac{2}{3},\frac{4}{3};-\frac{\kappa ^3}{27}\right)\\ 
   &+\kappa ^2 \frac{\pi  \Gamma \left(\frac{2}{3}\right)}{6 \sqrt{3} \Gamma \left(\frac{1}{3}\right)^2} \, _3F_2\left(\frac{2}{3},\frac{2}{3},\frac{2}{3};\frac{4}{3},\frac{5}{3};-\frac{\kappa
   ^3}{27}\right)-\frac{\pi ^2}{9}.
\ea
\ee
In the analytic continuation of the orbifold prepotential we find a non-trivial integration constant, as in related examples \cite{dmp}:
\be
F_0(\kappa)= {4 \over 3} \zeta(3) -\frac{\pi ^2   \Gamma \left(\frac{1}{3}\right)}{\Gamma \left(-\frac{1}{3}\right)^2}\kappa +\CO(\kappa^2).
\ee
The generalized theta function transforms non-trivially under the flip of sign, since the $n$ in (\ref{tt}) gets shifted to $n+1/2$ 
(as in (\ref{theta12})), and the second factor appearing 
in (\ref{xi-orb}) is given by 
\be
 \Theta_{\rm orb}(\kappa, 2 \pi)= \sum_{n \in \IZ} \exp \left\{ {9 \pi \ri (n+1/2)^2 \over 4} \tau + 2 \pi \ri  (n+1/2)  \left( \xi+ {1\over 8} \right) - 3 \pi \ri 
 \left( n+1/2\right)^3  \right\},
\ee
where $\xi$, $\tau$ are now evaluated with the formulae (\ref{tau-f}) and (\ref{xi-f}), but 
using instead the standard prepotential. Like in previous cases, this function can be 
massaged into a conventional Jacobi theta function, and we find
\be
\Theta_{\rm orb}(\kappa, 2 \pi)= \re^{ \pi \ri/8} \vartheta_2 \left(\xi -{1\over 4}, \tau_{\rm orb} \right), 
\ee
where we have denoted 
\be
\tau_{\rm orb}= {9 \tau \over 4}- {1\over 2}. 
\ee
All the quantities involved in (\ref{xi-orb}) have now a well-defined expansion around $\kappa=0$, and we can proceed as in the related calculations 
in \cite{cgm}. However, there is a difference: at the orbifold point, $\tau_{\rm orb}$ is a cubic root of unity, and the expansion involves 
the values of the Jacobi theta function $ \vartheta_2$ and its derivatives at 
\be
\tau_{\rm orb}(\kappa=0)=\re^{2 \pi \ri \over 3}, \qquad \xi(\kappa=0)={1\over 12}. 
\ee
We are not aware of closed form expressions for these quantities, but by consistency with our results we find for example 
\be
\ba
\vartheta_2 \left({1\over 6} ,\re^{2 \pi \ri \over 3}\right)&=3^{-7/24}\frac{\sqrt{\Gamma \left(\frac{1}{3}\right)}}{\Gamma \left(\frac{2}{3}\right)} \re^{-\pi \ri/8} , \\
 \partial_\tau \vartheta_2 \left({1\over 6} ,\re^{2 \pi \ri \over 3}\right)&=\frac{\ri \sqrt{\Gamma \left(\frac{1}{3}\right)} \left(\pi ^{3/2}+12 \Gamma \left(\frac{7}{6}\right)^3\right)}{2 
   \pi ^{3/2}3^{19/24} \Gamma \left(\frac{2}{3}\right)} \re^{- \pi \ri/8}, 
   \ea
   \ee
 which can be checked numerically to high precision. After taking into account these and similar identities, one finds the expected result, 
\be
\Xi(-\kappa, 2 \pi)= 1- {1\over 9} \kappa + \left( {1\over 12 {\sqrt{3}} \pi}- {1\over 81} \right) \kappa^2+ \cdots.
\ee

We conclude that the orbifold theory of local $\IP^2$ contains information about the spectral traces of the operator (\ref{lp2-op}), as in the related calculation 
of \cite{cgm}. It is also remarkable that the complicated numbers involved in the expansion of each of the factors in (\ref{xi-orb}) finally combine into the simpler types of quantities appearing in 
the traces and the canonical partition functions 
$Z(N, 2\pi)$. This seems to be the number theory counterpart of the underlying analytic simplicity of the spectral determinant, as compared to its factors. It might also lead to interesting experimental 
identities for the values of the Jacobi theta function and its derivatives. 

\sectiono{Other examples}

In this section, we will present more evidence for our conjecture by analyzing two further examples: local $\IF_1$ and local $\IP^1 \times \IP^1$. In both cases, 
the spectrum of the relevant quantum operators has been extensively studied numerically in \cite{hw}, and this makes it possible to check our analytic results on the spectrum. 
In addition, we present tests of the conjecture (\ref{spec-conj}). 

\subsection{Local $\IF_1$}

The quantum operator corresponding to local $\IF_1$ can be read from Table \ref{table-ops}. This model has one parameter $m$, and for simplicity we will restrict ourselves to 
the case in which $t_m=0$. This corresponds to setting $m=1$, so that
\be
\widehat \CO(\hat x, \hat p)=\re^{\hat{x}}+\re^{-\hat{x}}+\re^{\hat{p}}+\re^{\hat{x}-\hat{p}}.
\ee
The semiclassical analysis of the grand potential of this operator is technically more involved than in other cases, although it can be deduced from the 
calculation of the classical volume of phase space by using the results of \cite{km}. Let us then proceed directly with the calculation 
of the quantum, modified grand potential. 

In the case of local $\IF_1$, $r=1$, and the $B$ field is $B=1$. The special geometry of the model is encoded in its 
Picard--Fuchs operator. In the current case 
it is given by the third order differential operator \cite{hkp}
\be
\ba
\mathcal{L}&=(-12m^2+9\tilde{u}-18m \tilde{u}^2+8m^2 \tilde{u}^3)\pd_{\tilde{u}}^3
+(-108m-128m^4+144m^2 \tilde{u}+27\tilde{u}^2\\
&\quad-64m^3\tilde{u}^2-52m\tilde{u}^3+24m^2\tilde{u}^4)\pd_{\tilde{u}}^2
+(-9+8m\tilde{u})(-27+16m^3+36m\tilde{u}\\
&\quad-8m^2\tilde{u}^2-\tilde{u}^3+m\tilde{u}^4)\pd_{\tilde{u}},
\ea
\ee
and we can solve for the periods in the case of interest $m=1$. As usual, there is a trivial solution $\varpi_0(z)=1$, and 
\be
\ba
\varpi_1(z)&=\log(z)+\widetilde{\varpi}_1(z),\\
\varpi_2(z)&=\log^2(z)+2\widetilde{\varpi}_1(z)\log(z)+\widetilde{\varpi}_2(z),
\ea
\ee
where $\widetilde{\varpi}_1(z)$ and $\widetilde{\varpi}_2(z)$ can be easily found order by order in an expansion around $z=0$, 
\be
\ba
\widetilde{\varpi}_1(z)&=z^{2}-2z^{3}+\frac{3}{2}z^{4}-12z^{5}+\frac{55}{6}z^{6}+\cdots,\\
\widetilde{\varpi}_2(z)&=-\frac{1}{4}z+\frac{15}{16}z^{2}-\frac{91}{36}z^{3}
+\frac{231}{64}z^{4}-\frac{6403}{300}z^{5}+\frac{115}{3}z^{6}+\cdots.
\ea
\ee
The standard prepotential is determined by
\be
\ba
Q&=\re^{-t}=z \exp\left( \widetilde{\varpi}_1(z)\right)=z+z^3-2 z^4+2 z^5-14 z^6+22 z^7+\cO(z^{8}),\\
\pd_t F_0(t)&=4\left(\log^2 z+2 \widetilde{\varpi}_1(z) \log z+\widetilde{\varpi}_2(z) \right). \\
\ea
\ee
Then we easily find, 
\be
\widehat F_0(t)=\frac{4}{3}t^3-Q-\frac{15 Q^2}{8}-\frac{82 Q^3}{27}-\frac{15 Q^4}{64}-\frac{626 Q^5}{125}-\frac{205 Q^6}{36}-\frac{2402 Q^7}{343} + \CO(Q^8), 
\ee
and we have 
\be
J_{\rm WS}(\mu, \hbar)= - \left( 2 \sin {2 \pi^2 \over \hbar} \right)^{-2} \re^{-2 \pi \mu/\hbar} +\cdots, 
\ee
The value of $C(\hbar)$ can be read from the value $C=4$. The value of $B(\hbar)$ has been computed in \cite{hw}, and we have
\be
C(\hbar)=\frac{2}{\pi \hbar}, \qquad B(\hbar)= \frac{\pi}{3\hbar}-\frac{\hbar}{12\pi}. 
\ee
By using the quantum mirror map of local $\IF_1$, which has been worked out in \cite{hkrs,hw}, we obtain
\be
\mu_\text{eff}=\mu-\re^{-2\mu}+\frac{1+q}{\sqrt{q}}\re^{-3\mu}-\frac{3}{2}\re^{-4\mu}+\frac{1+5q+5q^2+q^3}{q^{3/2}}\re^{-5\mu}+\cdots.
\ee
With these ingredients, we can already find the quantization condition. As for local $\IP^2$, we write it 
as (\ref{exact-qc}), where $\Omega(E)$ is given in (\ref{approx}), and the 
correction $\lambda(E)$ is determined by (\ref{qc-corr}). We find, for the very first orders, 
\be
\lambda(E)=\lambda_1\re^{-16 \pi E_{\rm eff}/ \hbar}+\lambda_2\re^{-18 \pi E_{\rm eff}/ \hbar} +\lambda_3\re^{-22 \pi E_{\rm eff}/ \hbar}+\mathcal{O}(\re^{-24 \pi E_{\rm eff}/ \hbar}),  
\ee
where
\be 
\ba \lambda_1=& \frac{1}{\pi } \sin (16 x),\\
  \lambda_2=&\frac{1}{\pi }\sin (18 x),\\
  \lambda_3=&\frac{2}{\pi }(\sin (16 x)+2 \sin (20 x)+\sin (24 x)).
 \ea 
 \ee
 The coefficient $\lambda_1$ agrees with the correction computed in \cite{hw} by numerical fitting. This already gives a non-trivial test for our conjecture in the case of local $\IF_1$, and at the 
 level of the spectrum.

\begin{figure}
\center
\includegraphics[height=5cm]{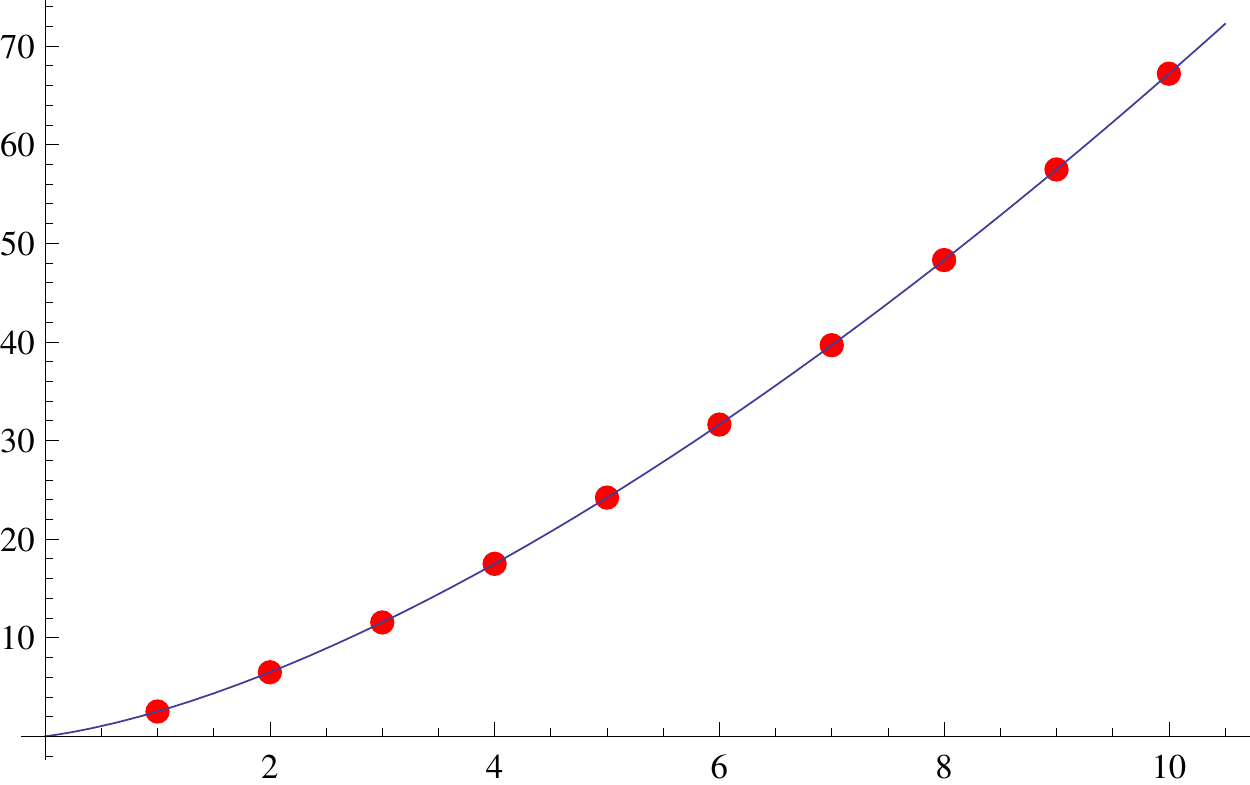}  
\caption{The smooth line gives the free energy $F(N, 2 \pi)$ of local $\IF_1$ as a function of $N$, computed from (\ref{zn-airy}), while the points give the values of 
of the free energy as computed from the spectral traces.}
\label{znf1}
\end{figure}

In order to test our conjecture for the spectral determinant, let us focus again on the supersymmetric case $\hbar=2\pi$. In this case, the effective chemical 
potential is given by 
\be
\mu_{\rm eff}= \mu-\widetilde \varpi_1(-z)=\mu-\re^{-2\mu}-2 \re^{-3\mu}-\frac{3}{2}\re^{-4\mu}-\cdots.
\ee
We can now use the general formula (\ref{mgp-2}) to write down the modified grand potential. As in the case of local $\IP^2$, the genus one free energies can be written in closed form 
(for $F_1$, see for example \cite{hkr}), and we find, 
\be
\ba
\widehat F_1(t)&=-\frac{1}{2}\log \left( \frac{\rd}{\rd z} \log Q \right)-\frac{1}{12}\log \Delta(z)-\frac{2}{3} \log z, \\
\widehat F_1^\text{NS}(t)&=-\frac{1}{24} \log \left( {\Delta(z)\over  z^4} \right),
\ea
\ee
where the discriminant $\Delta(z)$ is given by
\be
\Delta(z)=1+z-8z^2-36z^3-11z^4.
\ee
For $\hbar=2 \pi$, one has $B(2 \pi)=0$. From the general expression (\ref{mgp-2}) one can compute the large $\mu$ expansion of $J(\mu, 2 \pi)$:
\be
\ba
J(\mu,2\pi)&=\frac{\mu^3}{3\pi^2}+A(2\pi)-\( \frac{\mu^2}{8\pi^2}
+\frac{\mu}{4\pi^2}+\frac{1}{4\pi^2}+\frac{1}{8}\)\re^{-\mu} \\
&\quad-\( \frac{31}{16\pi^2}\mu^2+\frac{15}{16\pi^2}\mu+\frac{15}{32\pi^2}-\frac{1}{16} \)\re^{-2\mu} \\
&\quad-\( \frac{133}{24\pi^2}\mu^2+\frac{41}{18\pi^2}\mu+\frac{41}{54\pi^2}-\frac{11}{24} \)\re^{-3\mu}+\cdots.
\ea
\ee
The value of $A(2 \pi)$ can be computed numerically from the condition (\ref{normz}), and it reads
\be
A(2\pi) \approx 0.3075779653...
\ee
In this case it is more difficult to fit it to an exact expression. This might be related to the fact that the 
value of the coordinate $z$ at the conifold point has a non-trivial expression, since it is the solution to an algebraic 
equation of degree four. From all this information, we find the spectral determinant as a particular case of (\ref{ms-xi}), 
\be
\label{sd-f1}
\Xi(\mu,2\pi)=\re^{J(\mu,2\pi)} \vartheta_3\left( \xi+{1\over 3},{\tau \over 4} \right).
\ee
The quantization condition when $\hbar=2 \pi$ is given by the specialization of (\ref{qc-max}) to our case. Note that, now, 
\be
E_{\rm eff}= E-\widetilde \varpi_1\left(\re^{-E}\right). 
\ee
From this quantization condition we can compute the spectrum with very high precision. For example, for the ground state energy, we find 
\be
E_0=2.8640042594081906825951812...
\ee
to be compared to the numerical result obtained by diagonalizing (\ref{omatrix}) with $L=300$, 
\be
E_0^{(300)}=2.8640042594...
\ee
We can then compute the spectral traces numerically from the quantization condition, and from them the values of the canonical partition functions $Z(N, 2 \pi)$. 
We find again that the results obtained in this way agree with the prediction of (\ref{spec-conj}), i.e. with the formula (\ref{zn-airy}), as shown in \figref{znf1}. We could in principle 
obtain the spectral traces from the expansion of (\ref{sd-f1}) around the orbifold point $z=\infty$, as we did in the case of local $\IP^2$.

\subsection{Local $\IP^1 \times \IP^1$}

The case of local $\IP^1 \times \IP^1$ is known to be closely related ABJ(M) theory, whose spectral determinant was analyzed in \cite{cgm} in the 
maximally supersymmetric cases. The general ABJ(M) theory 
will be the object of a separate publication \cite{ghm-abjm}. We will then summarize the most important points of this case, since some of our results can be taken {\it verbatim} from \cite{cgm,ghm-abjm}. 

The quantum operator can be read from Table \ref{table-ops}. This model has one parameter $m$, and for simplicity we will restrict ourselves again to 
the case in which $t_m=0$. This corresponds to setting $m=1$, and the operator reads
\be
\widehat \CO(\hat x, \hat p)=\re^{\hat{x}}+\re^{-\hat{x}}+\re^{\hat{p}}+\re^{-\hat{p}}.
\ee
This operator has an additional interest, since it corresponds to the simplest case of the quantum, relativistic Toda lattice (see for example \cite{sergeev}). The results we will find have then 
implications for this integrable system.

As in the case of local $\IP^2$, it is easy to work out the semiclassical limit of the grand potential, which turns out to be identical to the case of ABJM theory 
studied in \cite{mp}. The classical spectral traces are given by 
 \be
Z_\ell^{(0)}
 =\frac{1}{2\pi}\frac{\Gamma^4(\ell/2)}{\Gamma^2(\ell)}, 
\end{equation}
which lead to 
\be\mathcal{J}_0(\mu)=-\frac{\kappa ^2} {8 \pi } \, _4F_3\left(1,1,1,1;\frac{3}{2},\frac{3}{2},2;\frac{\kappa ^2}{16}\right)-\frac{\pi  \kappa}{4}   \, _3F_2\left(\frac{1}{2},\frac{1}{2},\frac{1}{2};1,\frac{3}{2};\frac{\kappa ^2}{16}\right).
\ee
Expanding around $\kappa= \infty$ 
we find
\be \mathcal{J}_0(\mu)=\frac{2 \mu ^3}{3 \pi }+\frac{\pi  \mu }{3}+\frac{2 \zeta (3)}{\pi }+
 \left( -\frac{4 \mu ^2}{\pi }+\frac{4 \mu }{\pi }-\frac{2 \pi }{3}+\frac{4}{\pi }\right) \re^{-2\mu}+ \mathcal{O}(\re^{-4\mu}).\ee
It follows that
\be C(\hbar)= {2 \over \pi \hbar}, \quad B(\hbar)= { \pi \over 3 \hbar}+ \mathcal{O}(\hbar). 
\ee
Let us now obtain the results for the modified grand potential at finite $\hbar$. In this case we have that $r=2$. The large radius periods are given by
\be
\ba 
 \widetilde \varpi_1 (z)&=\sum _{n\geq 1}16^n \frac{ 1}{n}\left(\frac{\Gamma \left(n+\frac{1}{2}\right)}{\Gamma \left(\frac{1}{2}\right) n!}\right)^2z^n ,\\
 \widetilde \varpi_2 (z)&=
  \sum _{n\geq 1} \frac{4}{n} 16^n \left(\frac{\Gamma \left(n+\frac{1}{2}\right)}{\Gamma \left(\frac{1}{2}\right) n!}\right)^2\left( -\frac{1}{2 n}-\psi (n+1)+\psi \left(n+\frac{1}{2}\right)+\log (4) \right)z^n. 
 \ea 
 \ee
 The prepotential is then given by
 \be 
  F_0^{\rm inst}(t)=-4 Q-\frac{9 Q^2}{2}-\frac{328 Q^3}{27}-\frac{777 Q^4}{16}+\mathcal{O}(Q^5),
   \ee
where $Q=\re^{-t}$ is obtained by the mirror map
\be 
Q=z+4 z^2+\mathcal{O}(z^3). 
\ee
The $B$ field is in this case $B=2$ and has no effect on the signs. The worldsheet instanton piece of the grand potential, $J_{\rm WS}(\mu, \hbar)$, reads
\be 
J_{\rm WS}(\mu, \hbar)=-4 \left( 2 \sin {2 \pi^2 \over \hbar}\right)^{-2} \re^{-4 \pi  \mu / \hbar} +\cdots .
\ee
By using the quantum mirror map for local $\mathbb{P}^1 \times \mathbb{P}^1$ \cite{acdkv}, one has
\be  
\pi \hbar J_a (\mu)=-4 \re^{-2\mu} +\left(-2 q-\frac{2}{q}-14\right)\re^{-4\mu}   +\left(-4 q^2-\frac{4}{q^2}-24 q-\frac{24}{q}-\frac{232}{3}\right)  \re^{-6\mu}+\cdots,
\ee
where $q$ is again given by (\ref{q-def}). The coefficients appearing in (\ref{jper}) are given by
\be
 C(\hbar)=\frac{2}{\pi  \hbar },\quad B(\hbar)=\frac{\pi }{3 \hbar }-\frac{\hbar }{12 \pi }, \ee
 where the quantum correction to $B(\hbar)$ is found in \cite{hw}. 
 As in local $\mathbb{P}^2$, the constant term is related to the constant map contribution
 \be
 A(\hbar)=\frac{3}{2}A_\text{c}\( \frac{\hbar}{\pi} \)-A_\text{c}\( \frac{2\hbar}{\pi} \).
\ee
 As in the previous cases, 
 we can write down the quantization condition from the generalized theta function. If we write
 \be 
 \lambda(E)= \lambda_1\re^{-16 \pi E_{\rm eff}/ \hbar}+ \lambda_2\re^{-20 \pi E_{\rm eff}/ \hbar}+
  \lambda_3\re^{-24 \pi E_{\rm eff}/ \hbar}+\lambda_4\re^{-28 \pi E_{\rm eff}/ \hbar}+ \mathcal{O}(\re^{-32 \pi E_{\rm eff}/ \hbar}),
 \ee
we find from (\ref{qc-corr}) 
\be \ba \lambda_1&=  \frac{1}{\pi }\sin (16 x), \\
\lambda_2&={4 \over \pi}\csc ^2(2 x)\left(4 \sin ^2(4 x) \sin (20 x)\right),\\
\lambda_3 &= {8\over \pi} \csc ^2(2 x)\left(3 \sin ^2(4 x) \sin ^2(6 x)+\sin ^2(2 x) \sin ^2(8 x)+\sin ^2(10 x)\right) \sin (24 x), \\
\lambda_4 &={8\over \pi}\csc ^2(2 x)\sin ^3(4 x)\Big(45+52 \cos (4 x)+90 \cos (8 x)+52 \cos (12 x)+87 \cos (16 x)\\
&+48 \cos (20 x)+68 \cos (24 x)+26 \cos (28 x)+22 \cos (32 x)+4 \cos (36 x)+3 \cos (40 x) \Big),\\
\ea
\ee
where $x$ is given again by (\ref{xvar}). The first three coefficients reproduce precisely the corrections to the approximate quantization condition (\ref{approxWKB}) found in \cite{hw} by numerical 
fitting. 

In order to test our conjecture (\ref{spec-conj}), let us focus again on the maximally supersymmetric case with $\hbar=2 \pi$. It turns out that, in this case, the spectral problem 
becomes identical to the one in ABJ theory with $k=2$ and $M=1$. This is due to the fact that the partition function of 
ABJ theory with level $k$ and flux $M$ can be also described by topological string theory on local $\IP^1 \times \IP^1$ \cite{mori-abj,ho-abj}. The value of the parameter $m$ appearing in the 
operator of table \ref{table-ops} is related to $k$ and $M$ by 
\be
m =\exp \left( \ri \pi k - 2 \ri \pi M \right). 
\ee
Therefore, for $k=2$, $M=1$, we have $m=1$, as in the case we are considering here. It is easy to confirm that the quantization condition for $\hbar=2 \pi$, 
\be 
\label{susiv}
4 E_{\rm eff}^2-{4 \pi^2 \over 3}-2 \partial_t F_0^{\rm inst}(t)+4 E_{\rm eff}\partial_t^2 F_0^{\rm inst}(t)  = 4 \pi^2 \left(s +{1\over 2}\right) , \quad s=0,1,2, \cdots 
\ee
where 
\be 
E_{\rm eff}=E-2 z \, _4F_3\left(1,1,\frac{3}{2},\frac{3}{2};2,2,2;16 z\right), 
\ee
can be written as
\be
 \label{cqNf2}
 \ba
& 4\pi  {K(1-16 \re^{-2E})  \over K(16 \re^{-2E} )} \left(E- 2 \re^{-2E} \, _4F_3\left(1,1,\frac{3}{2},\frac{3}{2};2,2,2;16 \re^{-2E} \right) \right) \\
& \qquad  - {2\over  \pi} G_{3,3}^{3,2}\left(16 \re^{-2E} \left|
\begin{array}{c}
 \frac{1}{2},\frac{1}{2},1 \\
 0,0,0 \\
\end{array}
\right.\right)=4\pi^2 \left(s +1/2 \right), \qquad s=0,1,2,\cdots
\ea
\ee
In this equation, $K(k^2)$ is the elliptic integral of the first kind. This is precisely the quantization condition found in \cite{cgm} for 
ABJ theory with $M=1$ and $k=2$ (in this maximally supersymmetric case, it agrees with the approximate 
quantization condition (\ref{approxWKB}), which was studied in \cite{kallen}). This condition leads for example to a ground state energy 
\be 
E_0=2.881815429926296782477...
\ee

The analysis of the spectral determinant is in this case almost identical to what was done in \cite{cgm}. In particular, the traces, as computed from the energy spectrum, are given by 
\be \label{traceNf} 
\ba
Z_1&=\frac{1}{4 \pi }, \\
Z_2&= \frac{12-\pi ^2}{64 \pi ^2},\\
Z_3&= \frac{12-\pi ^2}{384 \pi ^3}, \\
Z_4&= \frac{96+80 \pi ^2-9 \pi ^4}{9216 \pi ^4}, \\
\ea
\ee
and so on. We will now verify that we can reproduce this result from (\ref{spec-conj}). 
We have to compute first the modified grand potential. For $\hbar=2 \pi$, the effective chemical potential is given by
\be 
\mu_{\rm eff}= \mu -{1\over 2}  \widetilde \varpi_1 (\re^{-2 \mu}).
\ee
The standard genus one free energy is given by
\be 
F_1(t)=-\frac{1}{2} \log \left(\frac{K(16 z)}{\pi }\right)-\frac{1}{12} \log (64 z (1-16 z)), 
\ee
while the refined genus one free energy is given by \cite{hk}
\be
 F_1^{ \rm NS}(t)=-\frac{1}{24} \log { 1-16 z \over z^2}.
 \ee
By using that
\be 
C(2 \pi)=\frac{1}{\pi ^2}, \quad B(2 \pi)=0,
 \ee
we get\footnote{Notice that the convention for $F_0$ is the same as in \cite{cgm},  except for the fact that here we do not include the constant $A(2\pi)$ in it. Moreover, in \cite{cgm} $t=2 \mu_{\rm eff}+\ri \pi$, while here $t=2 \mu_{\rm eff}$.}
\be \label{JNf}J(\mu,2 \pi)= A(2\pi)+{1\over 4 \pi^2}\left( F_0(t) - t \partial_t F_0(t) +{ t^2\over 2} \partial_t^2 F_0(t) \right)+ F_1(t)+F_1^{\rm NS}(t),
\ee
where 
\be  t= 2 \mu_{\rm eff},
\ee
and the constant $A(2 \pi)$ is given by \cite{hatsuda-o}
\be 
A(2 \pi)=\frac{1}{2} \left(\log (2)-\frac{\zeta (3)}{\pi ^2}\right). 
\ee
We then find, for the large $\mu$ expansion of the grand potential, 
\be 
\ba J(\mu,2 \pi)=&\frac{\mu ^3}{3 \pi ^2}+\frac{1}{2} \left(\log (2)-\frac{\zeta (3)}{\pi ^2}\right)-\frac{\left(4 \mu ^2+2 \mu+ 1\right) }{\pi ^2}\re^{-2\mu}
\\
& +\left( -\frac{52 \mu ^2+\mu +\frac{9}{4}}{2 \pi ^2}+2 \right)\re^{-4\mu}+\mathcal{O}({\re^{-6\mu}}), 
\ea
\ee
which agrees with the result for ABJ theory with $k=2$, $M=1$ \cite{ho-abj,hatsuda-o}. The spectral determinant is given 
by\footnote{The parameter $\tau$ is as in \cite{cgm}, while the $\xi$ variable 
there is given in terms of our $\xi$ variable by $\xi+{1\over 4}+{\tau \over 4}$. This is due to the difference in the definition of $t$.}
\be 
\Xi(\kappa, 2 \pi)=\exp\left( J(\mu, 2 \pi)\right) \vartheta_3\left( \xi-{1 \over 3 }, \tau\right). 
\ee
In order to reproduce the spectral traces (\ref{traceNf}), we have to expand this determinant around $z=\infty$ or $\kappa=0$. As in the case of local $\IP^2$, this means expanding the 
spectral determinant around the orbifold point of local $\IP^1 \times \IP^1$. To do this, we have to write it in terms of orbifold quantities. One has
\be 
\Xi(\kappa, 2 \pi)= \exp\left( J_{\rm or}(\kappa, 2 \pi) \right)\vartheta_1\left(\bar \xi +{1\over 2}, \bar \tau\right), 
\ee
where
\be 
\bar \tau=-{1 \over \tau}, \quad  \bar \xi= {\xi+{1\over 6} \over \tau}.\ee
In this formula the grand potential  $J_{\rm orb}(\kappa, 2 \pi)$ is given by (\ref{JNf}), where we subtract the constant pice $A(2 \pi)$ and we do the following replacements: we replace $F_1$ by 
\be 
F_1^{\rm orb}=- \log (\eta (2 \bar \tau)),  
\ee
we replace $t$ by
\be 
\lambda={\ri \kappa \over 8 \pi}  {~}_3F_2\left(\frac{1}{2},\frac{1}{2},\frac{1}{2};1,\frac{3}{2};\frac{\kappa^2
   }{16}\right), 
   \ee
and the the genus zero free energy $F_0$ should be replaced by $F_0^{\rm orb}$, where  
\be
\label{dFor} \partial_\lambda F_0^{\rm orb} (\lambda)=
 -{ \ri \kappa } G^{2,3}_{3,3} \left( \begin{array}{ccc} {1\over 2}, & {1\over 2},& {1\over 2} \\ 0, & 0,&-{1\over 2} \end{array} \biggl| {\kappa^2\over 16}\right)
 +{ 2\pi^2  \kappa }   {~}_3F_2\left(\frac{1}{2},\frac{1}{2},\frac{1}{2};1,\frac{3}{2};\frac{\kappa^2
   }{16}\right). \ee
The integration constant is fixed by requiring 
   \be F_0^{\rm orb}= 16 \pi^2 \lambda^2 \left( \log (2 \pi \lambda)-{3\over 4}- \log 4\right)+\cdots, \quad  \lambda \ll 1.\ee
   A simple computation shows that
   \be 
   \ba &\bar \xi=-{1\over 4}- {\ri \over 16 \pi^3} \left( \lambda\partial_\lambda^2 F_0^{\rm orb}-\partial_\lambda F_0^{\rm orb} \right) ,\\
   &\bar \tau = {1\over 32 \pi^2 \ri }\partial_\lambda^2 F_0^{\rm orb}. \ea\ee
   With the above ingredients we find
   \be
    \Xi (\kappa, 2 \pi)=1 +{\kappa \over 4 \pi}+\frac{\kappa^3}{128} \left(1-\frac{8}{\pi ^2}\right)+\cdots. \ee
 From this expansion we can read the canonical partition functions $Z(N, 2 \pi)$, which correspond indeed to the traces (\ref{traceNf}).

\sectiono{Conclusions and outlook}

In this paper, inspired by recent results on ABJM theory, and in particular by the work of \cite{hmmo,km}, 
we have proposed a correspondence between the spectral theory of functional difference operators, 
and the enumerative geometry of toric CY manifolds. This proposal leads to concrete and testable conjectures 
on the spectral properties of the operators obtained by quantization of mirror curves. 
One important point in our proposal is that the NS limit of the refined topological string is {\it not} enough to give a consistent description of the spectrum, 
as it was already pointed out in \cite{km}: non-perturbative effects involving the conventional topological string have to be added. The quantization condition proposed in 
\cite{km} can be now understood as a
first approximation to the full quantization condition obtained in this paper, which matches all known results for the spectrum and in particular the beautiful 
numerical study performed in \cite{hw}\footnote{In \cite{km}, the relationship between the grand potential and the quantization condition was analyzed by 
integrating the density of states. Although this is very useful from the point of view of the WKB expansion (see appendix A for example), 
it leads to many technical complications in the study of the non-perturbative sector. 
In \cite{km} it was assumed that there would be cancellations leading to a simple quantization condition, but these only occur in some cases, like in the maximally supersymmetric situations.}. 

Another important ingredient of our conjecture is that the quantization condition determining the spectrum 
arises as a consequence of a stronger result, namely, an explicit formula for the spectral determinant of the operators. 
We have tested this new ingredient in some examples. It should be emphasized that having explicit formulae for spectral determinants is rather rare. Even in elementary 
Quantum Mechanics, there are very few cases where this happens, besides the harmonic oscillator. For example, in the work of Voros and of Dorey and Tateo on 
polynomial potentials in QM \cite{voros,dt}, 
functional difference equations for spectral determinants have been shown to determine the spectrum uniquely, but 
one is far from having explicit formulae like the ones we are proposing. 
In this sense, our results provide a full new family of solvable models in spectral theory. 

From a physical point of view, our proposal can be understood as a microscopic description of the topological string in terms of a Fermi gas, as it was done in \cite{mp} for ABJM theory. 
This proposal has many appealing features as a non-perturbative description of topological string theory: it is based on a background independent 
object, it involves an M-theoretic version of the topological string free energy, it leads to the standard genus expansion in the 't Hooft approximation, and 
it suggests an underlying description in 
terms of M2 branes. 

Our work leads to many interesting problems. The first one is clearly: why such a conjecture should be true? From the point of view of spectral theory, 
what we are saying is that there are instanton 
corrections to the quantization condition (a well-known fact, see for example \cite{km} for a discussion and references), and that these corrections are determined by the 
conventional topological string. In the case of ABJM or local $\IP^1 \times \IP^1$, this is largely ``explained" by the fact that the canonical partition function $Z_X(N, \hbar)$ has an explicit matrix model 
representation, whose 't Hooft expansion involves the standard topological string \cite{mpabjm,akmv-cs}. 
Our conjecture is based on the idea that such a 
remarkable structure can not be peculiar to local $\IP^1 \times \IP^1$, and should be 
shared by all toric CY manifolds. The data seem to indicate that our conjecture is largely correct, but if so it is definitely begging for a deeper explanation, and eventually for a proof. 

In the meantime, we should check the conjecture in more detail. Already in the original realm of toric del Pezzo's considered here, many things should be clarified and new examples should be addressed. 
For example, it would be important to have closed formulae for the constant $A(\hbar)$, which so far we had to compute numerically or to guess. In ABJM theory there is a proposal for this function 
based on the resummation of the constant map contribution to the topological string free energies \cite{hanada}. It is clear that this resummation will be one of the ingredients in $A(\hbar)$, but it is likely that 
there are additional ingredients. 

Another important issue is to analyze the spectral determinant in detail away from the ``maximally supersymmetric" cases. This involves dealing with the all-genus topological 
string free energy, resummed in the way proposed by Gopakumar--Vafa. Although this can be computed systematically at large radius (by using for example the topological vertex \cite{akmv}) it is not clear 
what is its behavior at other special points of moduli space. This raises the issue whether there is a Gopakumar--Vafa reorganization of the topological string free energy at 
those points. We have some evidence that, in some circumstances, the all-genus expansion can be resummed into an explicit function of the moduli, which can then be 
studied at different points of the moduli space, and in particular near the orbifold point \cite{ghm-abjm}. In fact, one can reverse the logic and argue that, in order for our conjecture (\ref{spec-conj}) 
to work, and to be able to expand the spectral determinant around $\kappa=0$, a Gopakumar--Vafa resummation near the orbifold point must be possible. 

It would be also interesting to see if we can relax the reality and positivity conditions set on the parameters of the spectral problem. We have imposed such conditions 
in order to have self-adjoint operators with a positive, discrete spectrum, but it might be possible to extend our results to cases in which $\hbar$ and the complex moduli of the CY 
are not necessarily real. 

In this paper we have focused on the case of toric CYs whose mirror curve has genus one. It would be important to work out the details of the extension 
to higher genus. Many of the ingredients of this extension are relatively straightforward, so let us outline how this should work. 
First of all, we will have, instead of a single modulus $\tilde u$, $g$ different moduli $\tilde u_i$, $i=1, \cdots, g$. The mirror curve $W_X$ will now depend on the $\tilde u_i$. 
Our formalism can be generalized immediately to this situation, since 
the modified grand potential is still given by the non-perturbative free energy proposed in \cite{hmmo}. It will now depend on $g$ chemical potentials 
$\mu_i$, $i=1,\cdots, g$, corresponding to the moduli $\tilde u_i$, so we will write it 
as $J_X (\mu_i, \hbar)$. The grand canonical partition function is now given by 
\be
\label{xihg}
\Xi_X (\mu_i, \hbar)=\sum_{n_i \in \IZ} \exp\left( J_X(\mu_i+ 2 \pi \ri n_i, \hbar) \right), 
\ee
and will lead to a generalization of the Riemann--Siegel theta function. A generalized quantization condition can be obtained in 
a similar way, as the vanishing condition for this generalized theta function. There is therefore a natural extension of our conjecture to the 
case of general toric manifolds, but more work is needed to understand the higher genus case in detail. 

It is likely that some readers will wonder what is the relation between the non-perturbative effects unveiled here, 
and the resurgence approach, which has been 
recently applied to ABJM theory and topological string theory. Resurgence provides a general strategy for 
constructing formal trans-series, which complement the standard perturbative 
expansion by adding exponentially small effects. This leads to a 
multi-parameter family of asymptotic expansions, which should be then Borel resummed (in particular, 
resurgence does not determine a unique non-perturbative completion.). Our approach here is very different. As we emphasized at several places in this paper, 
we use the power of M-theory to resum the asymptotic genus expansion, so we are again in the realm of analytic functions: our modified 
grand potential has a region of analyticity, and our basic quantity -the spectral determinant- is in fact an entire function. At the same time, 
we do find non-perturbative corrections, and it might be possible to re-code them in the language of resurgence and of conventional instanton 
corrections (i.e., in the language of string perturbation theory plus D-brane corrections). We can for example regard the 
contribution of the refined topological string to the modified grand potential, as a non-perturbative correction to the conventional 
topological string free energy. One might then try to reproduce such a contribution by using a 
trans-series, constructed perhaps with the techniques proposed recently in \cite{cesv1,cesv2}. Some 
preliminary exploration of this issue was already made in \cite{gmz}, where the Borel resummation of the genus 
expansion of the topological string on local $\IP^1 \times \IP^1$ was 
compared to the corresponding modified grand potential, and it was concluded that they differ in a trans-series contribution. 

Finally, it would be important to understand the implications of our conjecture for the correspondence between 
quantum integrable systems and supersymmetric gauge theories 
put forward in \cite{ns}. It is well-known that $\CN=2$ gauge theories can be geometrically engineered as limits of topological string theory on 
certain CY geometries \cite{kkv}. In this way, we might 
be able to recover from our results, not only exact quantization conditions for the corresponding quantum integrable systems, but also 
explicit results for their spectral determinants, which are not covered by the original conjecture of \cite{ns}.

\section*{Acknowledgements}
We would like to thank Vincent Bouchard, Jie Gu, Masazumi Honda, Rinat Kashaev, Albrecht Klemm, Daniel Krefl, Sanefumi Moriyama, Tomoki Nosaka, 
Kazumi Okuyama, Jonas Reuter and Leon Takhtajan for useful discussions and correspondence. Thanks 
also to Johan  K\"all\'en for collaboration on the study of local $\IP^2$ in the fall of 2013. The work of A.G. and M.M. is supported in part by the Fonds National Suisse, 
subsidies 200021-156995 and 200020-141329, and by the NCCR 51NF40-141869 ``The Mathematics of Physics" (SwissMAP).

\appendix

\section{Semiclassical correction to the grand potential}\label{sec:J1}
Here we derive the result (\ref{eq:J1}).
We follow the procedure in \cite{km}.
In the semiclassical limit $\hbar \to 0$, the spectrum is determined by the usual WKB quantization condition.
In our set-up, this is nothing but the condition
\be
\Omega_\text{p}(E)=  s+{1\over 2}, \qquad s=0,1,2, \cdots,
\ee
where $\Omega_\text{p}(E)$ is given by (\ref{pnp}).\footnote{%
The function $\Omega(E)$ is related to the quantum volume in \cite{km}
by ${\rm vol}(E)=2\pi \hbar \Omega (E)$.}
This function has the semiclassical expansion
\be
\Omega_\text{p}(E)=\sum_{k=1}^\infty \hbar^{2k-1} \Omega_k(E).
\ee
For the local $\mathbb{P}^2$ case, the leading and the next-to-leading contributions $\Omega_0(E)$ and $\Omega_1(E)$ 
were computed in \cite{hw}
\be
\ba
2\pi \Omega_0(E)&=\frac{9E^2-\pi^2}{2}+9\sum_{n=1}^\infty \frac{(3n-1)!}{(n!)^3} [ \psi(3n)-\psi(n+1)-E ]\re^{-3n E}, \\
\Omega_1(E)&=-\frac{\Omega_0''(E)}{72}.
\ea
\label{eq:Omega-pert}
\ee
The energy also has the expansion around $\hbar=0$,
\be
E(s)=\sum_{k=0}^\infty \hbar^k E^{(k)}(s)=\log 3+\frac{\sqrt{3}(2s+1)}{6}\hbar+\cO(\hbar^2).
\ee
Following the argument in \cite{km}, we find
\be
\ba
\mathcal{J}_0(\mu)&=\int_{\log 3}^\infty \rd E \frac{\Omega_0(E)}{\re^{E-\mu}+1},\\
\mathcal{J}_1(\mu)&=\int_{\log 3}^\infty \rd E \frac{\Omega_1(E)}{\re^{E-\mu}+1} 
-\frac{1}{24\sqrt{3}(1+3\re^{-\mu})}.
\ea
\label{eq:J0J1-int}
\ee
One can numerically check that the expression of $\mathcal{J}_0(\mu)$ in (\ref{eq:J0J1-int}) precisely
reproduces the analytic result (\ref{eq:J0-exact}).
The analytic form of $\mathcal{J}_1(\mu)$ can be computed as follows.
Using (\ref{eq:Omega-pert}) and integration by parts, we obtain
\be
\ba
\int_{\log 3}^\infty \rd E \frac{\Omega_1(E)}{\re^{E-\mu}+1}
&=-\frac{1}{72} \int_{\log 3}^\infty \rd E \frac{\Omega_0''(E)}{\re^{E-\mu}+1} \\
&=\frac{1}{24\sqrt{3}(1+3\re^{-\mu})}
-\frac{1}{72} \pd_\mu^2 \(   \int_{\log 3}^\infty \rd E \frac{\Omega_0(E)}{\re^{E-\mu}+1} \),
\ea
\label{eq:J1-int-part}
\ee
where we have used $\Omega_0(\log 3)=0$, $\Omega_0'(\log 3)=\sqrt{3}$ and
\be
\pd_E \(\frac{1}{\re^{E-\mu}+1} \)=-\pd_\mu \( \frac{1}{\re^{E-\mu}+1}\).
\ee
From (\ref{eq:J0J1-int}) and (\ref{eq:J1-int-part}), we finally obtain (\ref{eq:J1}).

\end{document}